\documentclass[journal,onecolumn]{IEEEtran}

\usepackage{cite}
\usepackage{amsmath,amssymb,amsfonts}
\usepackage{algorithmic}
\usepackage{graphicx}

\usepackage{subcaption}

\usepackage{textcomp}
\usepackage{xcolor}
    \usepackage[acronym]{glossaries}

    \newglossaryentry{Crosscutting}{name={crosscutting}, description={Properties or areas of interest such as quality of service, energy consumption, location awareness, users' preferences, and security}}
     \newglossaryentry{heterogeneous}{name={heterogeneous}, description={A set of collaborated aspects (code fragments), that extend the application behaviour in several parts of the program and have an impact across the whole software system}}
     \newglossaryentry{Homogeneous}{name={homogeneous}, description={Applying the same code, that extend the application behaviour in several parts of the program}}

   \newglossaryentry{Context}{name={context}, description={Any information that is computationally accessible and upon which behavioural variations depend}}
     \newglossaryentry{Self-adaptive}{name={self-adaptive}, description={A self-adaptive application modifies its own structure and behaviour in response to changes in its operating environment}}

    \newglossaryentry{Self-* properties}{name={self-* properties}, description={The autonomic properties of a software, which includes (self-organising, self-healing, self-optimising and self- protecting)}}
   \newglossaryentry{Self-organising}{name={self-organising}, description={The capability of reconfiguring automatically and dynamically in response to changes by installing, updating, integrating, and composing/decomposing software entities}}
    \newglossaryentry{Self-healing}{name={self-healing}, description={The capability of discovering, diagnosing and reacting to disruptions. It can also anticipate potential problems, and accordingly take proper actions to prevent a failure}}
  \newglossaryentry{Self-optimising}{name={self-optimising}, description={The capability of managing performance and resource allocation in order to satisfy the requirements of different users. End-to-end response time, throughput, utilisation and workload are examples of important concerns related to this property}}   
 \newglossaryentry{Self-protecting}{name={self-protecting}, description={The capability of detecting security breaches, anticipating problems and recovering from their effects. It has two aspects, namely defending the system against malicious attacks, and anticipating problems and taking actions to avoid them or mitigate their effects}}  
  \newglossaryentry{Context-dependent}{name={context-dependent}, description={A context-dependent application adjusts its behaviour according to context conditions arising during execution}} 
    
    \newglossaryentry{Component Collaboration Architecture}{name={Component Collaboration Architecture}, description={how to model the structure and the behaviour of components at varying and mixed levels of granularity}}

\newglossaryentry{Context-awareness}{name={context-awareness}, description={The software system is aware of its context, which is its operational environment}} 

  \newglossaryentry{Context-aware}{name={context-aware application}, description={refer to a class of software systems that are able to monitor and detect context changes in the environment where they operate}}

\newglossaryentry{DESMET}{name={desmet}, description={A methodology for evaluating software engineering methods and tools}}

 \protect \newglossaryentry {Joinpoint}{name={joinpoint}, description={ A point in the control flow of a program. In aspect-oriented programming a set of join points is described as a pointcut. A join point is a specification of when, in the corresponding main program, the aspect code should be executed.}} 

 \newglossaryentry {Aspect}{name={aspect}, description={An aspect of a program is a feature linked to many other parts of the program, but which is not related to the program's primary function. An aspect crosscuts the program's core concerns, therefore violating its separation of concerns that tries to encapsulate unrelated functions}} 
  \newglossaryentry {Pointcut}{name={pointcut}, description={Is a set of join points. Whenever the program execution reaches one of the join points described in the pointcut, a piece of code associated with the pointcut (called advice) is executed. This allows a programmer to describe where and when additional code should be executed in addition to an already defined behavior.}} 
  \newglossaryentry {Advice}{name={advice}, description={Describes a class of functions which modify other functions when the latter are run; it is a certain function, method or procedure that is to be applied at a given join point of a program}} 
 
  \newglossaryentry {Separation of concerns}{name={separation of concerns}, description={Is the process of separating a computer program into distinct features that overlap in functionality as little as possible.}} 

\newacronym[\glsshortpluralkey=cas,\glslongpluralkey=contrived acronyms]{aca}{aca}{a contrived acronym}
\newacronym{cosd}{COSD}{Context Oriented Software Development}
\newacronym{MDA}{MDA}{Model Driven Architecture}
\newacronym{COP}{COP}{Context-Oriented Programming}
\newacronym{AOP}{AOP}{Aspect-Oriented Programming}
\newacronym{DAOP}{DAOP}{Dynamic Aspect Oriented Programming}
\newacronym{AOSD}{AOSD}{ Aspect Oriented Software Development}
\newacronym{MDD}{MDD}{ Model Driven Development}
\newacronym{CBSD}{CBSD}{ Component-based Software Development}

\newacronym{ADL}{ADL}{ Architecture Description language}
 \newacronym{COCOMO II}{COCOMO II}{Constructive Cost Model II}
  
 \newacronym{PIV}{PIV}{ Platform Independent View}
 \newacronym{CIV}{CIV}{ Computation Independent View}
 \newacronym{PSV}{PSV}{ Platform Specific View}
 
\newacronym{PIM}{PIM}{ Platform Independent Model}
 \newacronym{CIM}{CIM}{Computation Independent Model}
 \newacronym{PSM}{PSM}{Platform Specific Model}
   \newacronym{QoS}{QoS}{Quality of Services}
    \newacronym{OOP}{OOP}{Object Oriented Programming}

   \newacronym{ATAM}{ATAM}{Architecture Trade-off Analysis Method }
   \newacronym{UML}{UML}{Unified Modelling Language }
    \newacronym{OMG}{OMG}{Object Management Group}
     \newacronym{ECA}{ECA}{Enterprise Collaboration Architecture}
      \newacronym{EDOC}{EDOC}{Enterprise Distributed Object Computing}
 
      \newacronym{CCA}{CCA}{Component Collaboration Architecture}
       \newacronym{MOF}{MOF}{Meta Object Facility }
       \newacronym{DSL}{DSL}{Domain Specific Language}
       \newacronym{ATL}{ATL}{Atlas Transformation Language}
       \newacronym{EMF}{EMF}{Eclipse Modelling Framework}
        \newacronym{MUSIC}{MUSIC}{Mobile USers In Ubiquitous Computing}
        \newacronym{U-MUSIC}{U-MUSIC}{Unanticipated dynamic-adaptation for Mobile USers In Ubiquitous Computing}
  \newacronym{MADAM}{MADAM}{Mobility and ADaptation enAbling Middleware}
              \newacronym{OSGI}{OSGI}{Open Services Gateway initiative framework}
               \newacronym{MOSEL}{MOSEL}{MOdeling Specification and Evaluation Language}
 
                            \newacronym{PM}{PM}{Person-Months}
  
                            \newacronym{TDEV}{TDEV}{Time to Develop}
                            \newacronym{CAMEL}{CAMEL}{Context	Awareness ModEling Language}

                            \newacronym{A-MUSE}{A-MUSE}{Architectural Modeling for Service Enabling in Freeband}

                            \newacronym{CASA}{CASA}{Contract-based Adaptive Software Architecture}

                            \newacronym{POI}{POI}{Places Of Interest}
                            \newacronym{SLOC}{SLOC}{Source Lines Of Code}
 
                            \newacronym{ASLOC}{ASLOC}{Adapted source lines of code}
                             \newacronym{UFP}{UFP}{Unadjusted Function Points}
                             \newacronym{RFID}{RFID}{Radio Frequency IDentification}
                              \newacronym{QR-code}{QR-code}{Quick Response Code}
                                \newacronym{DP}{DP}{Decision Point}
                                \newacronym{DPL}{DPL}{Decision PoLicy}
                                \newacronym{CCTV}{CCTV}{Closed-Circuit TeleVision}
                                \newacronym{CAUCE}{CAUCE}{Context-aware Applications for Ubiquitous Computing Environments}
                                 \newacronym{AspectJ}{AspectJ}{Aspect-oriented Java extension} 
                                 \newacronym{ECORE}{ECORE}{Eclipse CORE meta-model}
\newacronym{JCOOL}{JCOOL}{Java COntext Oriented Language} 
\newacronym{JCOP}{JCOP}{Java Context-Oriented Programming} 
 \newacronym{Theme/UML}{Theme/UML}{Theme/Unified Modelling Language} 
 \newacronym{COCA-MDA}{COCA-MDA}{Context-Oriented Component-based Applications Model-Driven Architecture}

    \newacronym{COCA-middleware}{COCA-middleware}{Context-Oriented Component-based Applications Middleware}
 \newacronym{COCA-component}{COCA-component}{Context-Oriented Component model}
       \newacronym{COCA-ADL}{COCA-ADL}{Context-Oriented Component-based Applications Architecture Description Language}
       \newacronym{VML}{VML}{Virtual Machine Layer}
 
        \newacronym{COSM}{COSM}{Context-Oriented Software Middleware}
        
\begin{document}
 \title{Context Oriented Software Middleware}

\author{\IEEEauthorblockN{Basel Magableh}\\
\IEEEauthorblockA{\textit{School of Computer Science, } \\
\textit{Dublin Institute of Technology,}\\
\textit{Technological University}\\
Dublin, Ireland \\
basel.magableh@dit.ie}}
\maketitle

\begin{abstract}
Our middleware approach, \gls{COSM}, supports context-dependent software with self-adaptability and dependability in a mobile computing environment. The \gls{COSM}-middleware is a generic and platform-independent adaptation engine, which performs a runtime composition of the software's context-dependent behaviours based on the execution contexts. Our middleware distinguishes between the context-dependent and context-independent functionality of software systems. This enables the \gls{COSM}-middleware to adapt the application behaviour by composing a set of context-oriented components, that implement the context-dependent functionality of the software. Accordingly, the software dependability is achieved by considering the functionality of the COSM-middleware and the adaptation impact/costs. The COSM-middleware uses a dynamic policy-based engine to evaluate the adaptation outputs, and verify the fitness of the adaptation output with the application's objectives, goals and the architecture quality attributes. These capabilities are demonstrated through an empirical evaluation of a case study implementation.
\end{abstract}

 
 \section{Introduction}
 Mobile computing infrastructures make it possible for mobile users to run software systems in heterogeneous and resource-constrained platforms. Mobility, heterogeneity, and device limitations create a challenge for the development and deployment of mobile software. Mobility induces \gls{Context} changes to the computational environment and therefore changes to the availability of resources and services. This requires software systems to be able to adapt their functionality/behaviour to the context changes \cite{Inverardi:2009p2345}. This class of software systems are called \Gls{Context-dependent}/self-adaptive applications, which have the ability to modify their own structure and behaviour in response to context changes in the environment where they operate \cite{Oreizy:1999p3722}. Such level of self-adaptability presents the challenge of tailoring behavioural variations dynamically to both a specific user needs and adapt to the context changes. Moreover, because of the software pervasiveness, and in order to make adaptation effective and successful, adaptation processes must be considered in conjunction with software dependability and reliability by providing dynamic verification and validation mechanism, which validates the adaptation output with the adaptation goals, objectives, and architecture quality attributes \cite{Cheng:2009p3763,delemos:2011p3156}. Software dependability refers to the degree to which a software system or component is operational and accessible when required for use \cite{Barbacci:2010p4077}.

We believe that software self-adaptability and dependability can be achieved by dynamically composing software from context-oriented modules based on the context changes rather than composing the software from functional-oriented modules. Such composition requires the design of software modules to be more oriented towards the context information rather than being oriented towards the functionality. The principle of context orientation of software modules was proposed in the \gls{cosd}  \cite{magableh:2011p1231}. \gls{cosd} proposes a decomposition mechanism of software based on the separation between context-dependent and context-independent functionality. Context-independent functionality refers to software functionality whose implementation is unaffected by the context changes. For example, the map view and user login forms in mobile map application are context-free functionality (i.e. context changes would not change their functionality). The context-dependent functionality refers to software functionality, which exhibits volatile behaviour when the context changes. Separating the context-dependent functionality from the context-independent functionality enables software systems with adaptability and dependability with the aid of middleware technology. The middleware can adapt the software behaviour dynamically by composing that interchangeable context-dependent modules based on context changes. We argue that software self-adaptability is achieved by having both an adaptive middleware architecture and a suitable decomposition strategy, which separates the context-dependent functionality from the context-independent functionality of the software systems.

The objective of this research is to explore how far it is possible to support the development of self-adaptive applications using generic and platform independent middleware architecture provided by a non-specialised programming language such as \gls{COP}, and \gls{AOP}, and not limited to a specific platform or adaptation mechanism. This gives the software developers the flexibility to construct a self-adaptive application using generic reusable middleware components that employ popular design patterns and to implement their applications in several mobile computing platforms. 

Our \gls{COSM}-middleware is a generic and platform-independent adaptation engine, which performs a runtime behavioural composition of the context-dependent functionality of the software based on the operational context. The software dependability is achieved through evaluating the middleware functionality and the adaptation impact/costs. The \gls{COSM}-middleware uses a dynamic policy-based engine to evaluate and verify the fitness of the adaptation output among the application's objectives, goals and the architecture quality attributes. These capabilities are demonstrated through an empirical evaluation of a case study implementation. The implementation targets the construction of a self-adaptive map personalisation application, called eCampus. The objective of this evaluation is to measure the impact of the \gls{COSM}-middleware implementation over the allocated resources of a mobile device. The evaluation comes in two folds. The first experiment compares the performance and self-adaptability of the case study application implemented using \gls{COSM}-middleware and \gls{DAOP} engine. The second experiment compares the performance of the \gls{COSM}-middleware implementation with several frameworks and middleware architectures, including \gls{JCOP} \cite{JCOL:2011p22222}, \gls{JCOOL}\cite{Sindico:2009p3478}, \gls{MUSIC}\cite{Geihs:2011p232}, and \gls{MADAM}\cite{Mikalsen:2006p4052}. 

This article is organised as follows: Section \ref{sec:case} illustrates a case study implemented by \gls{COSM}-middleware.
Section \ref{sec:MWDesign} provides an overall description of the \gls{COSM}-middleware internal components. A detailed description of the \gls{COSM} implementation is provided in Section \ref{sec:platform}. The Context-Oriented Software Middleware is evaluated in terms of energy utilisation and adaptation time as discussed in Section \ref{sec:expr}.  
Section \ref{sec:RelatedWork} provides an overview of the related work.

\section{Case study: Self-adaptive Map Application}
\label{sec:case}
eCampus is a self-adaptive map application, which helps students, lecturers, staff members and visitors to explore the campus of the Technological University Dublin (TU).  In order for context-awareness capabilities to be made available, different aspects of spatial data need to be exploited including location, semantics, and time. The context-aware functionality offered in this application can be very useful to staff, students, visitors and the general public alike when navigating and otherwise interacting within our eCampus environment, specifically but also within any local environment generally while at home or on-site through web-based or smartphone connections. There are two types of users envisioned for this application: registered and non-registered users. A registered user would include anyone who wishes to log on (e.g. using their student/staff ID) and a non-registered user would include visitors to the campus (e.g. general public) who are not required to log on. As such, both user types are presented with different levels of functionality.
The registered users profiles are recorded so that their personal timetables and interests can be displayed. This can be displayed with a grid/table of all activities for the user on the current day (course, lectures during the day/week) associated with the venue. The other option is that the schedule is directly displayed on the personalised map where an overlaid vector layer identifies the geometry coordinates of the venue and detailed information about that activity. The recommendation of events is suited to each user profile (ie. based on their interests). The application selects the most related events to the user based on his profile. The features on the map refer to point-of-interest classified according to their score attribute ( calculated based on user's interests).

 In our case study, eCampus is required to adapt its behaviour and offer different levels of personalisation to the users depending on the available resources. For example, eCampus is required to adapt to the condition of low battery. This can be achieved by using a location service that consumes less power. The ``Battery level'' requirement needs a \gls{DPL} to manage its context changes and display less geospatial data to suit the battery level. In other words, the map application monitors the battery level and the bandwidth connectivity.  If the battery level is high and healthy, eCampus uses the GPS service with more accurate location's update, displays more features to the user, and animates the features' appearance on the map when they are recommended. If the battery level is less than a specific limit, a WIFI-based location is used to save the battery energy and obtain less geospatial data. Finally, if the battery level is low, the eCampus application switches off the GPS or WIFI location services and uses the cell-tower location services. Using a cell-tower location reduces the accuracy of the location but saves battery energy. In addition, the application may reduce the number of  features it displays on the map based on different interest score levels (values).


   \section{Context-Oriented Middleware}
   \label{sec:MWDesign}
\begin{figure}[!ht]
 \centering
     \includegraphics[scale=0.165]{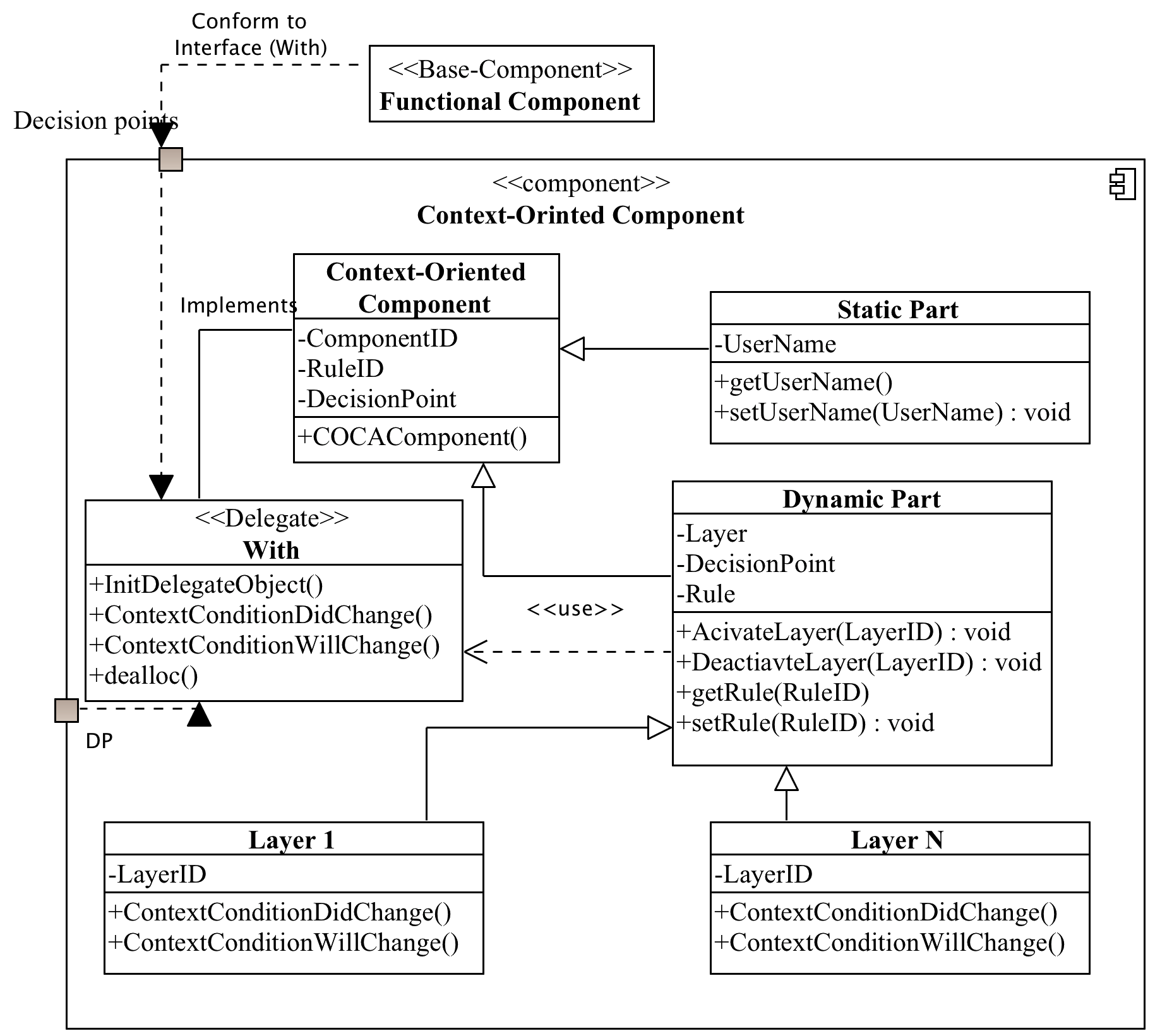}
    \caption{Context-Oriented Component Model} 
    \label{fig_com}
 \end{figure}

Before demonstrating the design of the context-oriented middleware, this section provides a general overview of the \gls{cosd} methodology. According to the \gls{cosd} methodology, context-oriented software is built from a set of components that follows the context-oriented component model \cite{magableh:2011p1231}. Figure \ref{fig_com} shows a conceptual diagram of context-oriented component. A context-oriented component consists of three major parts: static and dynamic parts, and a delegate object. The static part implements a context-independent code fragment responsible for implementing the application core functionality. The dynamic part implements the context-dependent functionality and participates in the context-driven adaptation. The dynamic part consists of multiple layers. Each component layer implements a specific context-dependent functionality, which implements a decision policy to be used for adapting specific application behaviour. A layer is executed only if the associated context condition is found in the environment at runtime. For example, small display, low memory, and low battery are contextual conditions that need to be considered in the adaptation. 

The \gls{COSM}-middleware  invokes a specific layer in the execution that suits only the current condition of execution context. For example, the \gls{COSM}-middleware needs to reduce the amount of geospatial data to consider low battery condition in a mobile map application. A decision-policy is implemented inside a specific layer method. The method is associated with the low battery condition. This method will be executed whenever the  \gls{COSM}-middleware receives the notification \textit{BatteryLevelWillChange}. In this case, the \gls{COSM}-middleware executes the layer method, which implements a decision-policy that displays less amount of geospatial data for the user (for example, displaying the features that have interest score between "0.7" and "1"). This approach can personalise the map content based on the availability of computational resources.   

The context-oriented component is given an opportunity to do dynamic context-driven adaptation by executing different layers, which implement different decision policies and methods implementation. Each layer must implement two or more methods that encapsulate decision-policies and associate them with context conditions. For example, the two methods inside the layer class of the context-oriented component, could have the names \textit{ContextConditionDidChange} and \textit{ContextConditionWillChange}. This allows the context-oriented component to perform adaptation action about a specific context condition in active or proactive mode. Active adaptation refers to the ability of the \gls{COSM}-middleware  to execute a layer when the condition is currently found in the execution context. Proactive adaptation refers to the ability of the \gls{COSM}-middleware to execute a decision-policy that will handle a condition that can happen after a certain amount of time. An example of active adaptation is when the user's location changes. The associated layer will execute a piece of code, that displays the user's location and nearly point-of-interests. On the other hand, an example of proactive adaptation is when the user's mobile is capturing high speed and acceleration. Our middleware can decrease the frequency of location-updates from the web mapping service until the speed of the mobile device is decreased to save allocated resources. More frequent updates of geospatial data is a power consuming process that needs more bandwidth and CPU throughput, consumes the allocated resources and decreases the battery life.

The fourth part of the context-oriented component is the delegate object see (Figure \ref{fig_com}). The idea of  using a delegate object is that two components coordinate to solve a problem. A context-oriented component is general and intended for reuse in a wide variety of contextual situations. The base-component is the digital map object, which stores a reference to that context-oriented component (i.e. its delegate) and sends messages to inform the delegate (context-oriented component) that some context condition was changed. This gives the delegate an opportunity to de/activate a layer implementation (i.e. rule implementation) dynamically \cite{buck:2010p4162}.

Our map application case study is designed from a set of base-components (the context-independent components) and context-oriented components. Whenever the map application notifies the \gls{COSM}-middleware about changes on context, the middleware activates a layer implementation that adapts to the changes by specifying an adaptation action that suits the context condition and considers the allocated resources. The following describes the structure of context-oriented software and the internal components of the \gls{COSM}-middleware.

 \begin{figure}[!ht]
\centering
\includegraphics[scale=0.6]{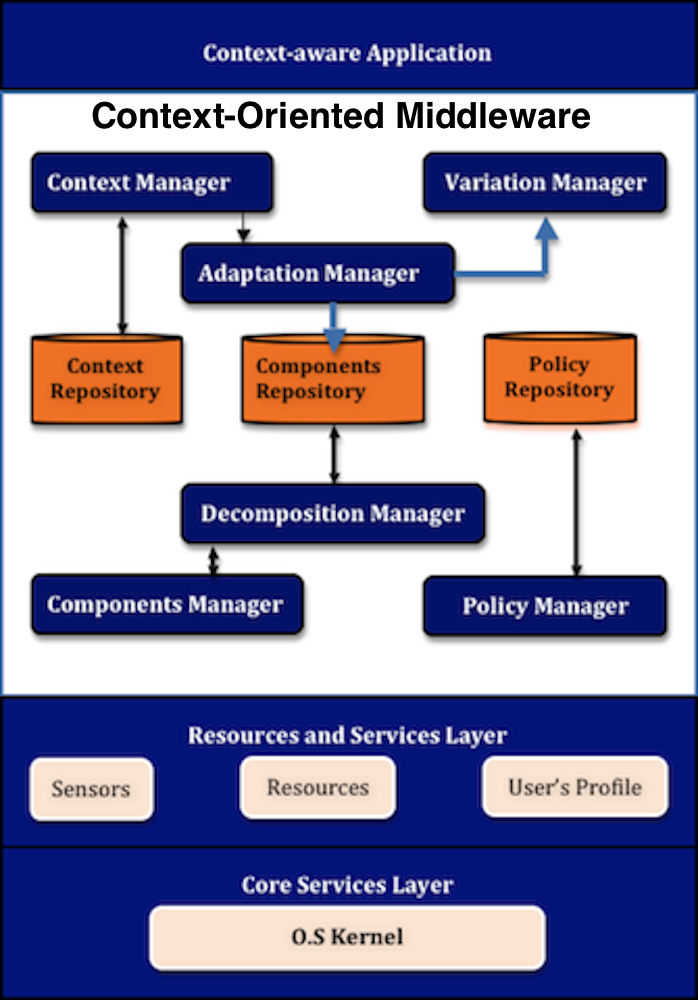}
\caption{Context-Oriented Software Architecture.}
\label{fig_platform}
\end{figure}

 In this article, we explore the design practices that can be used to implement the middleware architecture without relying on a specific framework for performing dynamic adaptation in context-dependent software systems. To address this issues, this section focuses on describing the \gls{COSM}-middleware architecture. The result of COSD  methodology is a component-based architecture described by a \gls{COCA-ADL}. The final step of the \gls{cosd} is to generate the application code and the \gls{COCA-ADL} XML file. The \gls{COCA-ADL} provides a description of the components, connectors, and the architecture's configuration, COCA-ADL is a platform-independent model that can be transformed by a model-to-model transformation tool into the desired platform-specific model. This provides code mobility for the same application into various deployment platforms and provides a runtime model of the self-adaptive application. However, the use of COSD for building self-adaptive applications for indoor wayfinding for individuals with cognitive impairments was proposed in \cite{magableh:2011p0002}. Evaluating the COSD productivity among the development cost and effort using \gls{COCOMO II} \cite{Boehm:2008p4159} was demonstrated in \cite{magableh:2011p2}. This article focuses on demonstrating the design principles and implementation of the \gls{COSM}-middleware, supported by an empirical evaluation of its performance and adaptability.

The \gls{COSM}-middleware offers a self-adaptive application a runtime support for adjusting the application's behaviour dynamically. Figure \ref{fig_platform} shows the context-oriented software architecture. The platform is layered into four major layers. Each layer provides an abstraction of the underlying technology. Each layer is platform independent of any given technology. The first layer represents the context-aware application. It provides the user with GUI, functional properties, and non-functional properties (i.e. the map view). The second layer in the platform represents the \gls{COSM}-middleware. The \gls{COSM}-middleware subcomponents are shown in Figure \ref{fig_platform}. The OS sensor retrieves information about the operating systems. Function calls are used to retrieve information about CPU, memory, and disk space \cite{Magableh:2009p2994}. The third layer represents the resources and services available in the execution platform, and the core services layer found in the O.S kernel.

 \subsection{Context Manager}
The first component of the \gls{COSM}-middleware is the context manager, as shown in Figure \ref{fig_platform}. The context manager gathers and detects context information from the sensors. If the context is changed, the context manager notifies the adaptation manager and the observers (i.e. the context-oriented components) about the changes. Each context-oriented component is designed to be an observer for one or more context entities. This type of interaction is called context binding. 

\begin{figure}[!ht]
  \begin{center}
    \includegraphics[scale=0.3]{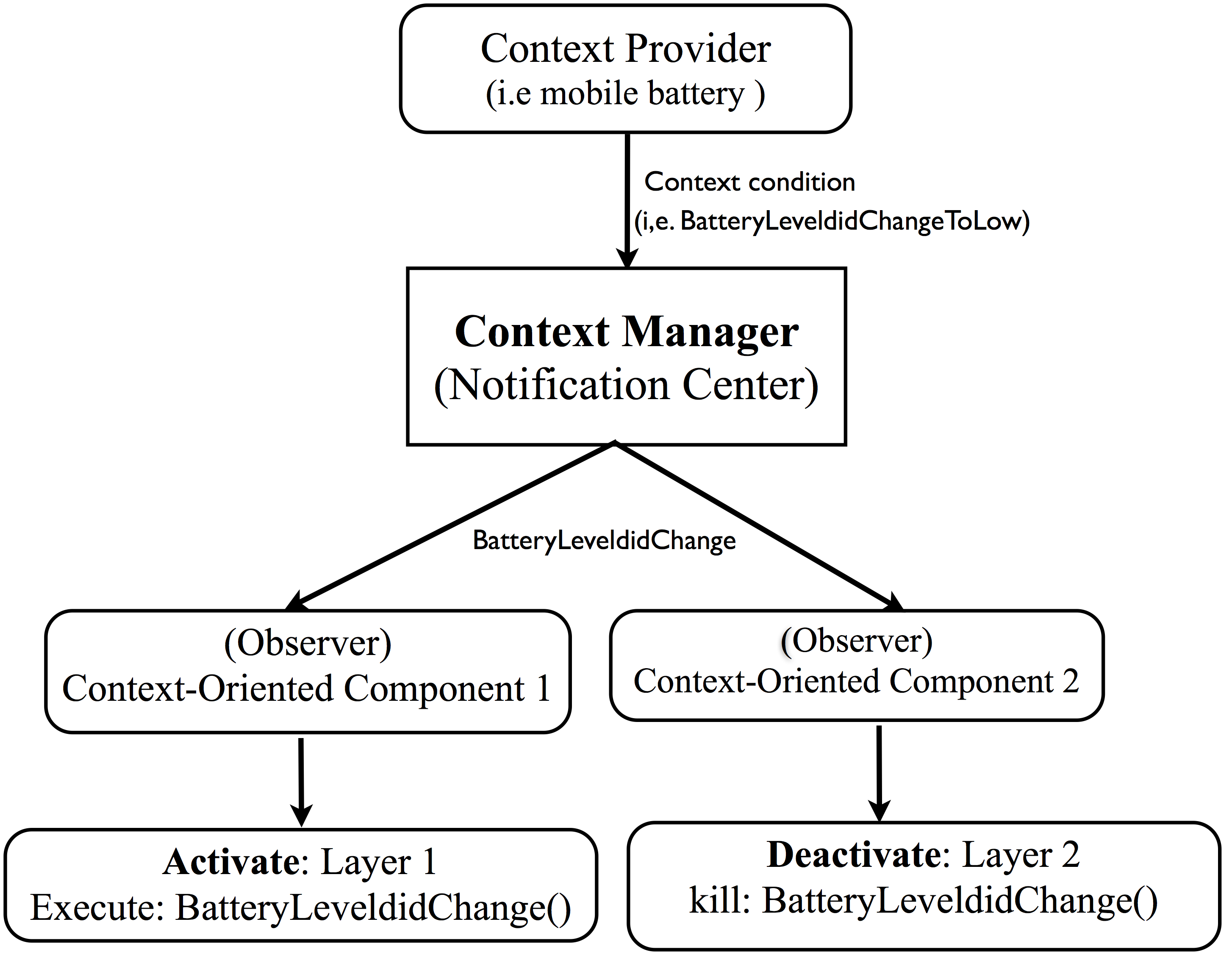}
    \caption{ Observer Design Pattern} \label{fig_observ} 
  \end{center}
\end{figure}
The observer pattern reduces the tight coupling between the context provider ( e.g. battery and memory) and the context consumer, (e.g. the context-oriented component). In addition, it enables the middleware to identify which context-oriented component has to be manipulated in response to context changes. Figure \ref{fig_observ} demonstrates the observer design pattern with one context entity and two observers. At runtime, both context-oriented components register themselves as observers for the context entity (e.g. \textit{BatteryLevelDidChange}) by sending a registration request to the notification centre. The context-change event is sent to the notification centre queue instead of the context-oriented component, then the notification centre broadcasts the context changes, and only the registered component receives the notification. The context-oriented components 1 and 2 have registered as an observer for the context entity. Whenever the context changes, both components 1 and 2 are notified by the notification centre and execute their embedded layer that may implements the methods (e.g. \textit{BatteryLevelDidChange}). In this way, the adaptation manager can identify  both components 1 and 2 to be included in the adaptation action, which embeds a subdivision of their implementation by de/activating the associated sub-layer. For example, the adaptation manager can activate layer 1 and executes the method (\textit{BatteryLeveldidChange()}) implemented in the first component, and deactivate layer 2 in the second component, which kills the execution of the method (\textit{BatteryLeveldidChange()}) as shown in the Figure \ref{fig_observ}. 

Supporting context-binding mechanisms with observer pattern provides a clear separation between the context provider and consumer. In addition, it classifies the components implementation based on the execution context conditions. This makes identifying which component must respond to a specific context condition an easy task in the design phase and runtime. However, to achieve this integration, the developers have to consider the following two aspects in the application design: how to notify the adaptation manager about context changes by identifying which components need to observe a specific context condition, and how the adaptation manager can identify the internal parts of the component that need to respond to these changes by clearly stating which method implements the needed behaviour for a specific condition.  
\subsection{Component Manager}
\label{co-com-imp}
 
The component manager performs three major functions in the middleware: it searches for a context-oriented component in the component repository, adds components from the repository, and provides context-oriented component instantiation. The component manager achieves the intercession operation by adding a component, or a component sub-layer, to the application structure. To add a component, the adaptation manager asks the component manager to instantiate a specific component. The component manager performs several inspections of the application components through the operation time of the software.  Further details about the implementation of the component manager are provided in Section \ref{co-com-imp}.

 \subsection{Policy Manager \label{sec:dp}}
 \glsreset{DPL}
 
 The \gls{cosd} provides the developers with the ability to specify the adaptation goals, actions, and causes associated with several context conditions using a policy-based framework. For each context-oriented component, the developers can embed one or more \glspl{DPL} that specify the architecture properties including (quality attributes, constraints and resources limits). At the design phase, the software developers use a state-machine model to describe the DPL by specifying a set of internal and external variables and conditional rules. The rules determine the true action or else an action based on the variable values. The action part of the state diagrams usually involves invoking one or more of the component's layers. A single layer is activated if a specific context condition is found, or deactivated if the condition is not found \cite{Anthony:2010p3204}. The policy manager uses the \gls{DPL} objects to store policies in the policy repository. The \gls{DPL} is stored in the policy repository, which conforms to the Associative Storage design pattern \cite{buck:2010p4162}. This pattern organises the policies into data and keys; this reduces the computation overhead from processing them at runtime.
\subsection{Decomposition Manager}
The decomposition manager operates as an XML parser for the COCA-ADL, and a constructor that constructs the application's components graph. At development time, the application's models are transformed into a \gls{COCA-ADL} XML file.  At runtime, the decomposition manager reads the \gls{COCA-ADL} XML file, and adds the architecture instances, including the components, connectors, and configuration, to the application graph. In other words, the decomposition managers constructs a runtime model of the software architecture as it is described by the COCA-ADL. This runtime model is manipulated by the adaptation manager during the adaptation process and It will describe the new adapted architecture of the software. 

\subsection{Adaptation Manager}
\glsreset{DPL}
The adaptation manager starts the adaptation process after receiving the notifications that identified the context changes and the context-oriented components that observed the notification. The first function of the adaptation manager is to produce the composition plan. The composition plan recursively describes the composite components and the connections between them by describing several connectors and interfaces. To construct a composition plan for \underline{ the new adapted software architecture}, the following information is needed by the adaptation manager. 1) A component graph: The decomposition manager generates the component graph after parsing the COCA-ADL XML file.  2) \gls{DPL}: The \gls{DPL} rules determine the true action or the else action based on the values of its variables. 3) Runtime structure style: When several context conditions are found at the same time, the \gls{DPL} proposes a runtime instance of a design pattern, which may imply combination of multiple components or their internal parts to fulfil the execution context. The structure style describes a structure modification combining a set of the component's layers. 

The adaptation process starts the adaptation action, including two types of composition mechanism: internal composition and external composition. In internal composition, the adaptation manager switches a component's layers on or off, based on the composition plan, using the delegation and decorator patterns. The decorator pattern can be used to extend, decorate, the functionality of a certain object at run-time, independently of other instances of the same component. The decorator pattern is an alternative to subclassing. Subclassing adds behaviour at compile time, and the change affects all instances of the original class; decorating can provide new behaviour at run-time for individual objects. In internal composition, the adaptation manager introspects the application's graph and achieves the self-tuning attribute of the self-adaptive software, self-tuning is achieved by redirecting the execution into the delegate object, which activates  the desired sub-layer implementation. 

In external composition, the adaptation manger adds or replaces components from the application structure, based on the composition plan. The decomposition manager builds the application graph by reading the \gls{COCA-ADL}. In external composition, a context-oriented component is loaded into the application. This requires the adaptation manager to conform to the bundle pattern \cite{buck:2010p4162}. The bundle pattern achieves the following goals: 1) Keep executable code and related resources together even when there are multiple versions and multiple files involved in the underlying storage. 2) Implement a flexible plug-in mechanism that enables dynamic loading of executable code and resources. In addition, the invocation design pattern is used to provide a means of capturing runtime messages so that they can be stored, rerouted, or treated and manipulated according to the context state, and allows new messages to be constructed and sent at runtime without requiring code recompilation process \cite{buck:2010p4162}. For example, when a component receives a message, a method implementation is usually invoked to handle the message. However, this is not always the case. As an illustration, if a component does not implement a particular method, then there is no method, which can be invoked and a runtime exception is raised instead. Because of the invocation design pattern, it is possible for a message to be delayed, rerouted to other components, or even ignored at runtime without re-compiling the application's code. 

In addition, the adaptor design pattern lets components work together, even if they have incompatible interfaces \cite{buck:2010p4162}. Assume a base-component needs to communicate with a context-oriented component, but its interfaces make that unachievable. To solve this problem, the context-oriented component applies to the delegate pattern by defining a protocol, which is essentially a series of method declarations unassociated with the component. The base-component then adapts the protocol and confirms this by implementing one or more of the protocol's methods. The protocol may have mandatory or optional methods. The base-component can then send a message to the protocol interface, which redirects the message to the context-oriented component. At this stage, the adaptation manager can verify whether the context-oriented component can respond to the message call, before invoking the message call by adapting the chain of responsibility pattern. This pattern verifies whether the component can respond to the method call using the responder pattern, which avoids coupling between the sender of a request and its receiver by giving more than one context-oriented component sub-layers a chance to handle the request. 
\subsection{Verification Manager}
 As long as the \gls{COSM}-middleware is aware of the architecture configuration, which is supported by the \gls{COCA-ADL} configuration element. The \gls{COSM}-middleware can anticipate the associated configuration with specific context changes. In each \gls{DP}, the \gls{COSM}-middleware transforms the software from $state_{i}$ into $state_{i+1}$, considering the properties of the self-adaptive software. These properties include the following: 1) The set of \glspl{DPL} attached to the context-oriented components that participate in the adaptation; 2) the architecture configuration elements in the \gls{COCA-ADL}, which includes the description of the \glspl{DPL} and the architecture properties, the \glspl{DPL} specify the external and internal variables that are evolving through the adaptation process; and 3) the adaptation goals, actions, and rules specified by the \glspl{DPL}.

\section{The Context-Oriented Software Middleware Runtime Platform}
 \label{sec:platform}
\begin{figure}[!ht] 
\centering 
\includegraphics[scale=0.5]{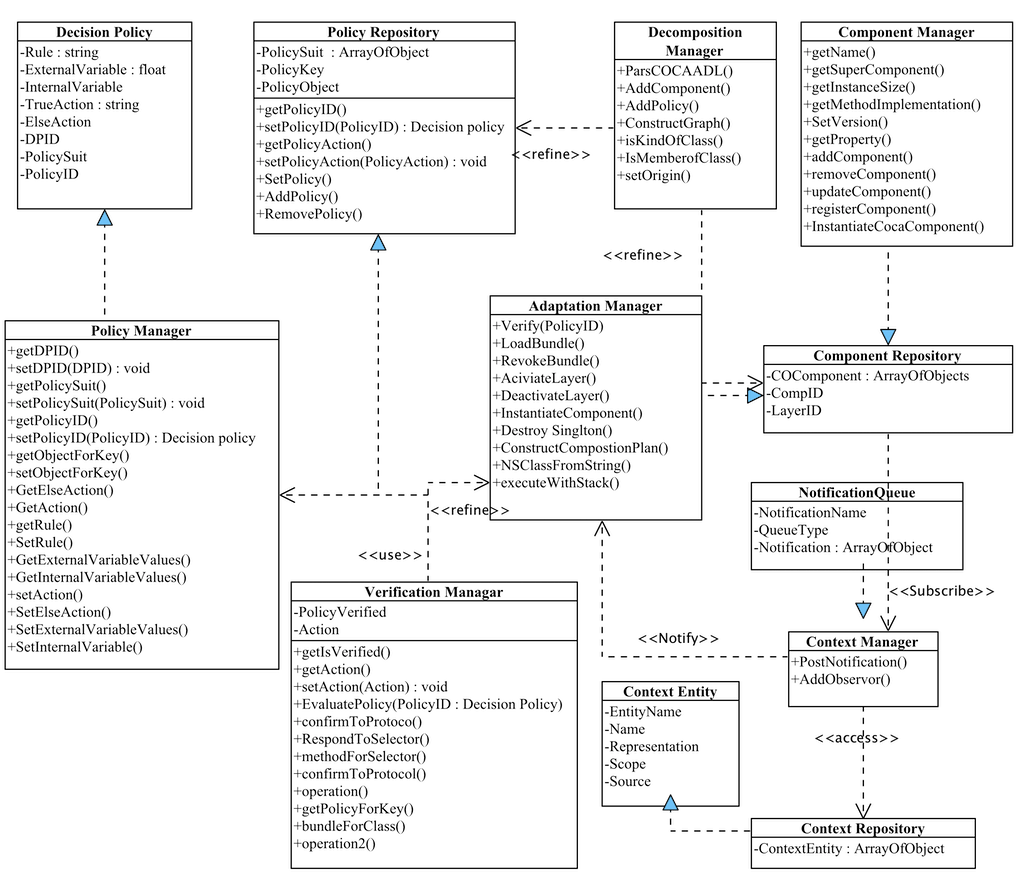}
\caption{The Context-Oriented Middleware Design} \label{fig_policyManager} 
\end{figure}

The final step of the \gls{cosd} is to generate the code and the \gls{COCA-ADL} XML file. The runtime functions start once these two inputs are in place. The major component in the \gls{COSM}-middleware is the adaptation manager. On a broader scale, the adaptation manager defers as many decisions as it can from compile time and link time to runtime. Whenever possible, it performs actions dynamically and executes the compiled code. Handling the context-oriented component framework, the composition plan, the application singleton, and the way in which several components interact with the runtime system is the focus of the following sections. Figure \ref{fig_policyManager} shows a class diagram for the \gls{COSM}-middleware. Both the  map application and \gls{COSM}-middleware were implemented in IOS mobile operating system  for IPhone devices \cite{IOS4:2011p3211}. For this reason, Objective-C is used to describe the method implementation of the \gls{COSM}-middleware components. 

The runtime platform is a dynamic shared library with a public interface, consisting of a set of functions and a data structure in the header file, located within the context-oriented framework. Many of these functions allow the application to perform the adaptation actions through the adaptation manager. The following sections describe the implementation of the \gls{COSM}-middleware components, each components is described in terms of the main design principles we used to implement it.

\begin{figure}[!ht] 
 \centering
 \includegraphics[scale=3.8]{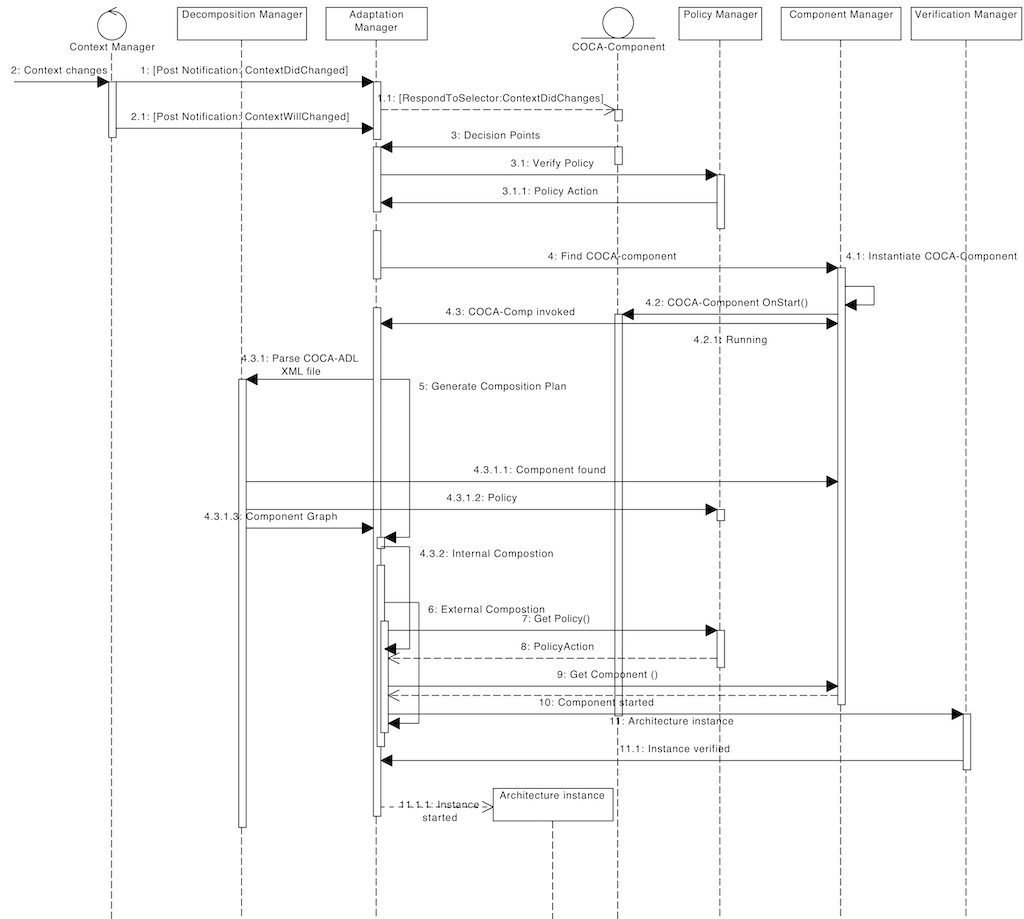} 
\caption{The Context-Oriented Middleware Adaptation Runtime Model } 
\label{fig_Runtime} 
\end{figure}

The context-oriented runtime platform starts once it has the compiled code for the application base-components and the context-oriented components plus the \gls{COCA-ADL} XML file. When the application is launched, the \gls{COSM}-middleware components are executed first. The adaptation manager then calls the decomposition manager to build the application composition graph and the inheritance tree. The decomposition manager parses the \gls{COCA-ADL} XML file for the component elements. The decomposition manager adds the components to the graph and the component repository. Each graph node has a component dispatch table. This table has entries that associate method selectors with the component-specific addresses of the methods they identify. In the same way, the decision policies are attached to the associated context-oriented component and added to the policy repository.  Figure \ref{fig_Runtime} shows a decomposition mechanism. Each component and its subdivisions are added to a graph in the buffer.

\subsection{Adaptation Manager Runtime Functions}

Once the decomposition is finished, the adaptation manager asks the context manager for the context state. At the same time, the adaptation manager runs the component instance for the base-component type. The adaptation manager checks each class by parsing the graph. In each node, the adaptation manager performs the following operations: 1) create the application singleton, 2) add the base-components to the singleton, and 3) construct the primary composition plan.

Afterwards, the context manager notifies the adaptation manager about the context state. Based on the context state, the adaptation manager reads the description of the component from the dispatch table. It confirms whether the component has the right objects, and methods which suit the context state. This is accomplished by asking the object to identify its class using \textit{isKindOfClass} and \textit{isMemberOfClass}. This verifies an object's position in the inheritance tree. The \gls{COSM}-middleware design conforms to the delegation design pattern. This requires each context-oriented component to define a protocol or a formal interface. During the composition, the adaptation manager identifies whether a specific component does conform to a protocol by calling  the method \textit{(conformsToProtocol:)}, which indicates whether an object claims to implement the methods in a specific protocol, then the operation \textit{(RespondToSelector:)} is performed, which indicates whether an object can accept a particular message. After that, the adaptation performs \textit{(methodForSelector:)}, which provides the address of a method's implementation. These methods enable the adaptation manager to introspect the application structure. The class diagram in Figure \ref{fig_policyManager} shows the relation between the adaptation manager class and the other \gls{COSM}-middleware components. 


 Parsing the \gls{COCA-ADL} XML file by the adaptation manager to construct the graph has some drawbacks with respect to device performance. A reasonable approach to parsing the XML file with less impact on the quality attributes is the use of the Flyweight design pattern \cite{buck:2010p4162}. The Flyweight pattern minimises the amount of memory and/or processor overheads required to use objects \cite{buck:2010p4162}. The Flyweight pattern enables instance sharing, to reduce the number of instances needed, while preserving the advantages of using objects. Classes that implement the Flyweight pattern are called ‘flyweights'. Flyweights encapsulate non-object data so that the data can be used in contexts where objects are required. Flyweights reduce storage requirements when a large number of instances are needed. Flyweights act as stand-ins for other objects \cite{buck:2010p4162}.

In addition to Flyweight, another pattern that can be used during implementation is the Associative Storage pattern. The most important feature of this pattern is the efficient storage of arbitrary data associated with objects; this promotes flexibility by delaying the selection of which data to access until runtime.

The \textit{MutableDictionary} data structure is used to implement the composition plan and the decision policies \cite{IOS4:2011p3211}. The use of \textit{XMLParser} implements an event-driven approach with a delegate object implementing methods for handling each of the `events' the parser encounters during its single pass over the XML data \cite{IOS4:2011p3211}. Events most commonly of interest are the beginning and ending of ADL elements and attribute data within elements. The \textit{XMLParser} reads the XML elements, then uses \textit{MutableDictionary} to store them in the dictionary. The \textit{setObject:ForKey:} method is used to create new associations in the dictionary. When keys and values are added and removed from a mutable dictonary, the memory allocated for storing objects grows and shrinks automatically. If \textit{setObject:ForKey:} is called with a key which is already in the dictionary, the object associated with that key is replaced by the new object. Each key is stored at most once \cite{IOS4:2011p3211}. 

After completing the composition plan, the adaptation manager implements the dynamic creation pattern to load and execute the application's components. Once the composition plan is completed, the adaptation manager introspects the application's structure. Redirecting the context-oriented component delegate to the desired layer activates the component sub-layers. In some cases, a context-oriented component is loaded into the application. This requires the adaptation manager to employ the Bundle and Invocation design patterns \cite{buck:2010p4162}. The \textit{forwardInvocation:} method is used to give a default response to the message, or to avoid the error in some other way. For example, suppose that the adaptation manager receives a message call for a method \textit{memoryLevelDidChange}. First the verification manager verifies whether the receiver object can respond to this message using \textit{respondToSelector} \cite{IOS4:2011p3211}. When the object cannot respond to the message because it does not have a method matching the selector message, the runtime system informs the object by sending it a \textit{(forward Invocation:)} message. Every object inherits  the method \textit{(forward Invocation:)} from the super-class context-oriented component.

 However, the object version of the method simply invokes \textit{doesNotRecogniseSelector}. In this case, the adaptation manager forwards to other objects. First, the adaptation manager determines where the message should go and sends the message with its original arguments. The message can be sent with the \textit{invokeWithTarget:} method. If the invocation has failed in the desired sub-layer, the method forwards the invocation to the context-oriented component, which forwards it into its sub-layers until one of the sub-layers responds to it. If the sub-layers do not respond to it, the adaptation manager introspects the component graph and reconstructs the composition plan. Once an object found in the distributed environment in a remote component. 
 
 The adaptation manager performs "invocation" by obtaining the method signature and the selector \cite{IOS4:2011p3211}. When a message is sent to an object that does not implement it, the actual implementation assumes that the stack frame for the arguments of the method already exists. All further changes to the method's arguments using Invocation's methods are performed on that stack frame. After the method invocation returns, the adaptation manager can access the returned value and possibly change it, using the \textit{(getReturnValue:)} and \textit{(setReturnValue:)} methods.

\subsection{Policy Manager Implementation}
As illustrated in Figure \ref{fig_policyManager}, the policy manager uses the decision policy objects to store policies in the policy repository. The policy dictionary stores each policy in loosely coupled objects, which are accessed through the method \textit{getObjectForKey}. Once the object is retrieved, the policy manager obtains the policy actions, attributes, rules, external variables, and internal variables. Then it passes them back into the verification manager. The verification manager evaluates the current value for the variables among the predefined values in the policy syntax. 

Whenever the application execution reaches decision points and/or the context manager has notified the adaptation manager of a context change, the decision points must be executed to advise the application of pre- or post-actions among specific notifications. The manager implements the necessary methods to manipulate the decision policy syntax. The policies are stored in an array of objects. Each object is accessed through the method \textit{getPolicyForKey}. The key refers to the policy ID that is attached to every context-oriented component. In the same way, the policy manager is used to upgrade the policy by calling the methods \textit{setObjectForKey}, \textit{SetPolicySuit}, \textit{SetRule}, \textit{setAction}, and \textit{setElseAction}.  

The decision policy is stored in the policy repository, which conforms to the Associative Storage design pattern. A binary representation for each policy is stored in the MutableDictionary data structure, where the key value is used to access the desired policy. The method \textit{addPolicy} is used to add a new policy to the repository. In the same way, policies can be removed using the method \textit{RemovePolicy}. Once the policy manager updates the policy, the method \textit{setPolicy} is used to update the policy syntax in the repository. The policy syntax is retrieved through the method \textit{getPolicyForKey}. This implementation of the Associative Storage design pattern reduces the computation overhead for retrieving the policies and evaluating them.

\subsection{Verification Manager Implementation}

A context-aware application is self-configurable if it is able to adapt autonomously to changing environmental conditions or internal status by altering its structures, behaviours, and data to meet its functionality and quality requirements. From a middleware perspective, such a feature relies on the following key characteristics: A) The middleware's ability to monitor and define its internal status and external conditions (e.g. application modes, CPU and memory use, and attachment of external devices); B) its built-in knowledge of configuration variability and related policies/rules for deciding and planning changes; and C) its ability to perform dynamic configuration changes without violating the constraints relating to overall system functionality, performance, and dependability. 

 The Flyweight pattern is used to guarantee that the verification process does not affect the quality attributes. This is accomplished by adapting the feature of instantiation in the Flyweight. Moreover, if there are many external and internal variables in the decision policy, all these variables will be instantiated once and share this instance to multiple values. In addition, the Flyweight acts as a temporary place holder for other more heavyweight objects. 

\begin{figure}[!ht] 
\centering 
    \includegraphics[scale=3.8]{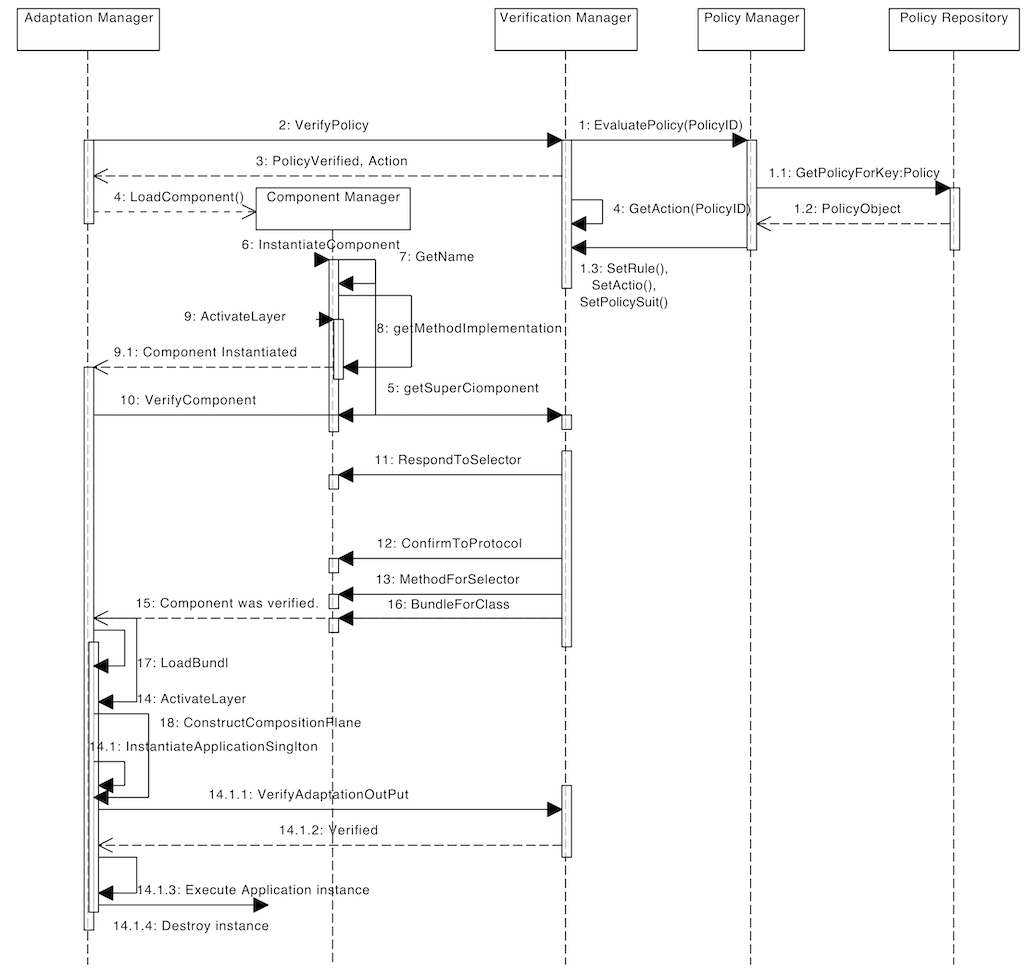}
    \caption{Adaptation self-assurance verification} \label{fig_assuranace} 
\end{figure}
Figure \ref{fig_assuranace} shows a sequence diagram of the self-assurance verification. The verification manager evaluates the policies by  calling the \textit{verifyPolicy:(NSInteger) PolicyID} method. This method asks the policy manager to retrieve the stored policy by its key. The policy manager searches the array of objects for the specific policy ID. The evaluation result, which contains the proposed action to be performed, is passed back to the adaptation manager. Afterwards, the adaptation manager locates the desired component and/or sub-layers which need to be executed. Then it asks the verification manager to verify them using the methods \textit{methodForSelector, confirmToProtocol, RespondToSelector}. Once the verification has been accomplished, the verification manager sets the boolean variable \textit{PolicyVerified} to be true.

The motivation behind the use of selectors is to postpone specifying the message that will be sent to an object until runtime. This reduces coupling between objects by limiting the information that message senders need about the message sent. In the same way, whenever the adaptation manager needs to verify the implementation of a context-oriented component or a subdivision, the selector will be used to determine whether  the object has the proposed method. The verification manager can load bundles/components by executing the method \textit{bundle = [NSBundle bundleWithPath:theBundlePath]}. However, after loading the bundle that contains a context-oriented component, the verification manager testifies its ability to respond to the desired method call using \textit{bundleForClass and respondToSelector} methods.


  \section{Context-Oriented Software Evaluation}
  \label{sec:expr}
  This section focuses on evaluating the performance and modifiability quality attributes of context-oriented software, including the \gls{COSM}-middleware and the case study implementation. Before evaluating the performance of the \gls{COSM}-middleware it is desirable to evaluate the modularisation strategy of context-oriented component to find out if there is any impact of such modular structure over the software self-adaptability and dependability. We do not know the impact of modularisation technique over software performance. The general belief is that using different decomposition mechanisms will not affect runtime properties of software. However, this may not apply to self-adaptive software systems. For this reason, the case study application was implemented using 
\gls{AOSD} \cite{asosd:2004p1972} and \gls{cosd}, as they adapt different decomposition techniques for supporting software with adaptability. Their ultimate goal is to support the adaptability and variability of software systems, and to be able to reduce development cost and effort, while improving the software modularity and complexity. This motivates us to evaluate these technologies with respect to their ability to support software adaptability (modifiability) and the performance gain from using these technologies to implement the case study application in a mobile computing environment. This article claims that \gls{cosd} and the \gls{COSM}-middleware are better suited to build self-adaptive software in the mobile computing domain. To this end, an evaluation of the two major paradigms (\gls{AOSD} and \gls{cosd}) is required to find out which one is better suited for building self-adaptive applications. 
\subsection{Metrics}
In the first experiment, the case study application eCampus was implemented as a real iPhone application using the \gls{cosd} and \gls{AOSD} paradigms. The second experiments compared the performance of the \gls{COSM}-middleware implementation with several frameworks and middleware architectures including \gls{JCOP} \cite{JCOL:2011p22222}, \gls{JCOOL} \cite{Sindico:2009p3478}, \gls{MUSIC} \cite{Geihs:2011p232}, and \gls{MADAM} \cite{Mikalsen:2006p4052}. Then the following questions were analysed. First, how expensive is it to perform context monitoring? Secondly, what is the effect on the allocated resources after the software has performed context detection, particularly when multiple heterogeneous events are detected? Thirdly, what is the performance gain of activating and executing multiple and collaborated aspects in comparison with context-oriented components composition, in response to multiple context events arriving at the same time. The following experiments focus on evaluating each paradigm implementation of the eCampus application to support adaptability and dependability, based on the following criteria.
\begin{enumerate}

\item \textbf{Battery Usage.} This criterion evaluates the device's battery durability while running the eCampus application and performing the adaptation processes, including context monitoring, detecting, decision-making, and adaptation. The IPhone Energy Diagnostic tool provides a relative energy usage on a scale of 0 to 20. These values explain how expensive it is with respect to the battery life to run a specific process over the execution time.

\item \textbf{CPU Activity.} This criterion analyses the CPU activities and CPU time required for performing the adaptation processes, including context monitoring, detecting, decision-making, and adaptation. This includes the time required for components/aspects composition in response to multiple and heterogeneous context change events.
\item \textbf{Real Memory Allocation.} This criterion measures the amount of memory allocated by the application during the execution of particular functionalities, including context monitoring, context detection, and adaptation.
\item \textbf{Sleep/Wake.} This parameter captures the eCampus application's ability to adjust its activity while the device is running in sleep mode. Normally, if the application keeps running in the background and performs some kind of operation, for example, updating the current location while the device is in sleep mode, the allocated resources are intensively degraded. This feature evaluates the architecture's ability to adjust the application behaviour, while considering the interoperability between the middleware functionality and the allocated resources.
\item \textbf{Adaptation/reconfiguration time.} This criterion captures the required time for the application to adapt its structure and behaviour by adding, removing, or updating components/services. The adaptation time was measured  from the start to the end of the adaptation action. The reconfiguration time measures the time required by the middleware to load and execute the plug-in (bundle) implementation.

\end{enumerate}

\subsection{Hardware and Software Configuration}
The CPU activity, CPU time, real memory allocation, and energy usage are measured for performing each adaptation process separately. These values were measured using the energy diagnostics and activity-monitoring tools, which analyses the running application on the iPhone device \cite{IOS4:2011p3211}. The Energy Diagnostic tool was used to measure the battery while the device was not connected to an external power supply; after the experiment was finished, the data were imported from the iPhone and then analysed. For measuring the CPU activity, CPU time and the real allocated memory, the eCampus application was executed on the same IPhone devices for each paradigm/framework implementation separately. The instrument tool was executed on Macbook pro, which traces the data from the IPhone device using the activity-monitoring tool. The IPhone was connected wirelessly with the activity-monitoring tool. This allows the tool to capture more accurate data for the energy usage and the CPU activity with respect to each process under evaluation. To testify variation in the application behaviour, a simulator was included with each eCampus implementation. The simulator is used to allow the user to simulate specific context changes, which are used to test the application's ability to adapt the desired behaviour. For each adaptation process, the experiment was established as follows:

\textbf{Context monitoring.} The CPU activity, CPU time, real memory allocation, and energy usage are measured for performing context monitoring at two different time intervals. First, when the application "did Finish Launching With Options" and the UI views did loaded. Second, the simulator interface was used to trigger the context monitoring process, this operation excluded any events related to the application load time. In addition, it allowed us to estimate the time required to process 10 contextual events enqueued at the same time. The simulator generates these events generally in First In First Out (FIFO) order. This experiment was executed 200 times, then the variance and standard deviation were calculated for the above criteria.  

\textbf{Context detection.} The CPU activity, CPU time, real memory allocation, and energy usage are measured for performing context detection with the aid of the simulator. When the context detection button is pressed, the simulator generates and en-queue multiple events, which are executed in First In First Out (FIFO) order. This allows the experiments to evaluate the required time to process these context events, and the time required to perform the reasoning action (i.e. decision-making) by the application. This experiment was executed 200 times, then the variance and standard deviation were calculated for the above criteria.  

\textbf{ Adaptation time/re-configuration time.} The CPU activity, CPU time, real memory allocation, and energy usage are measured for performing the adaptation/re-configuration in two modes. In the first mode, the application was executed for the same period of time (five hours). Then, the adaptation time was measured once the adaptation action was started until finished, the CPU time was taken from the activity-monitoring tool. In the second mood, the simulator was used to generate multiple context events, that measures the application response with regrade to low battery context. This experiment was executed 200 times, then the variance and standard deviation for each value were calculated. This allows us to identify the positive error as described in the following sections. 
  \subsection{COSD Vs. AOSD Experiments }

The assumption made by the \gls{AOSD} community is that dynamic aspect weaving can be used to adjust the software behaviour dynamically, regardless of the complexity involved in implementing Aspect-Oriented Programming (AOP) applications. Existing Dynamic AOP techniques tend to add a substantial overhead in both execution time and code size \cite{Hundt:2010p137}. The eCampus implementation was re-engineered to be integrated with the Objective-C AOP framework, called AspectCOCA proposed in \cite{AspectCOCA:2011}. The AspectCOCA framework offers many benefits to the AOP implementation, as it provides dynamic code weaving without any need to implement special AOP engine like PROSE 2 \cite{Popovici:2002p988}. In addition,  AspectCOCA implementation excluded the impact of Java virtual machine from the overall performance of the self-adaptive application, as AspectCOCA does not need a virtual machine to perform dynamic runtime operations including behavioural Intercession and Introspection. This offer great benefits for the application and reduces the development cost. We believe it is appropriate to implement the DAOP-eCampus  using this technology, which excludes the impact of aspect engine over the performance and device resources. As a result, we implemented several aspects for handling the context monitoring and detecting processes. In addition, the context-dependent behaviours for the location service, battery level, and the camera flashes were implemented.  However, for the location service, there are three nested aspects implemented to provide behavioural variation of the battery level. These aspects are the GPS-based, WiFi-based, and IP-based location services. In \gls{cosd}, these aspects are implemented using three context-oriented components.

  \subsubsection{Experiment 1: Context Monitoring and Sensing}
This experiment evaluates the processes of context monitoring and environment sensing, based on the above criteria. Specifically, it evaluates how the software uses the allocated resources such as battery, CPU, and memory. In the \gls{DAOP} approach, context monitoring is handled using  separate aspects which span the application's main execution. In \gls{cosd}, this is handled using the context manager.

Designing aspects that become active when particular contexts are verified requires the possibility of referring to a context definition in a \gls{Pointcut} construction. This means that \glspl{Joinpoint} such as BeInContext(Context BatteryLowCTX) should be provided by the framework. In addition, the \glspl{Aspect} composition needs to keep track of past context conditions and their associated states, which require more CPU activity and memory allocation to perform this functionality. This adds much overhead to the \gls{Advice} execution, because the \gls{AOP} framework must perform context snapshots through the monitoring and sensing process. The problem behind this is that the context snapshot is made every time the context is changed. This makes the platform storing and processing the context history for multiple events at multiple times. In addition, the \gls{AOP} framework must transform the context changes into basic entities like \gls{Joinpoint} request. The \glspl{Joinpoint} are activated by registering them to the execution monitor. When the execution reaches one of the activated \glspl{Joinpoint}, the execution monitor notifies the \gls{DAOP} engine, which executes the \gls{Advice} method. This implies, that the \gls{AOP} will evaluate each \gls{Joinpoint} with regard to the passive and active context through the context detection and decision-making processes as shown in the next experiment. 

 As a result, battery energy is consumed faster in comparison to the implementation based on \gls{COSM}-middleware, which includes a dedicated context manager supported by a context repository ( used to store and process the past context information). In addition, each context-oriented component registers its interest on a specific context change. This makes the context manager sense the environment for a particular set of context information.

\begin{figure}[!ht]
\centering
 \includegraphics[scale=0.54]{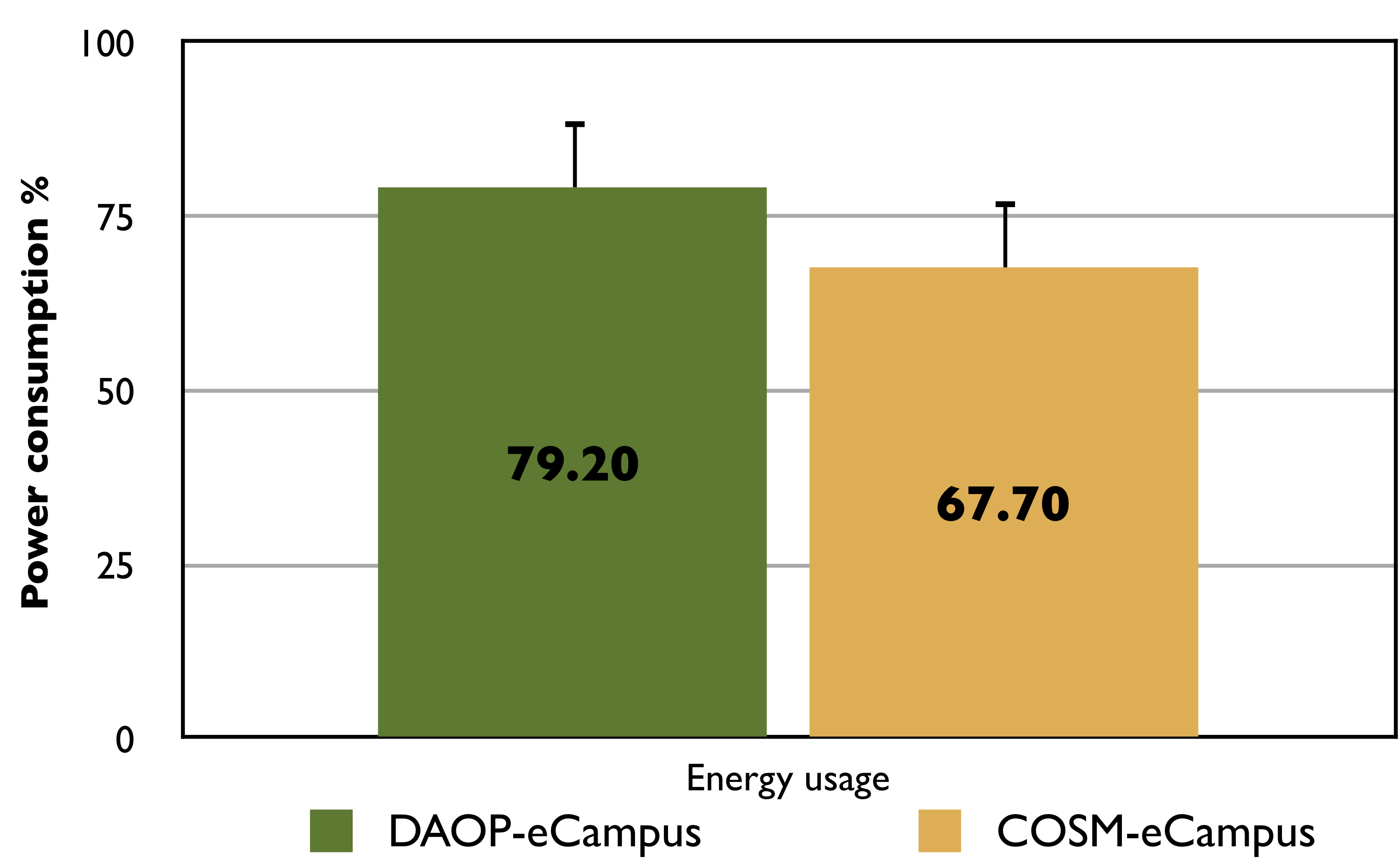}
  \caption{ Context Monitoring battery usage } 
\label{fig_conexp}
 \end{figure}

\begin{figure}[!ht]
\centering
 \includegraphics[scale=0.54]{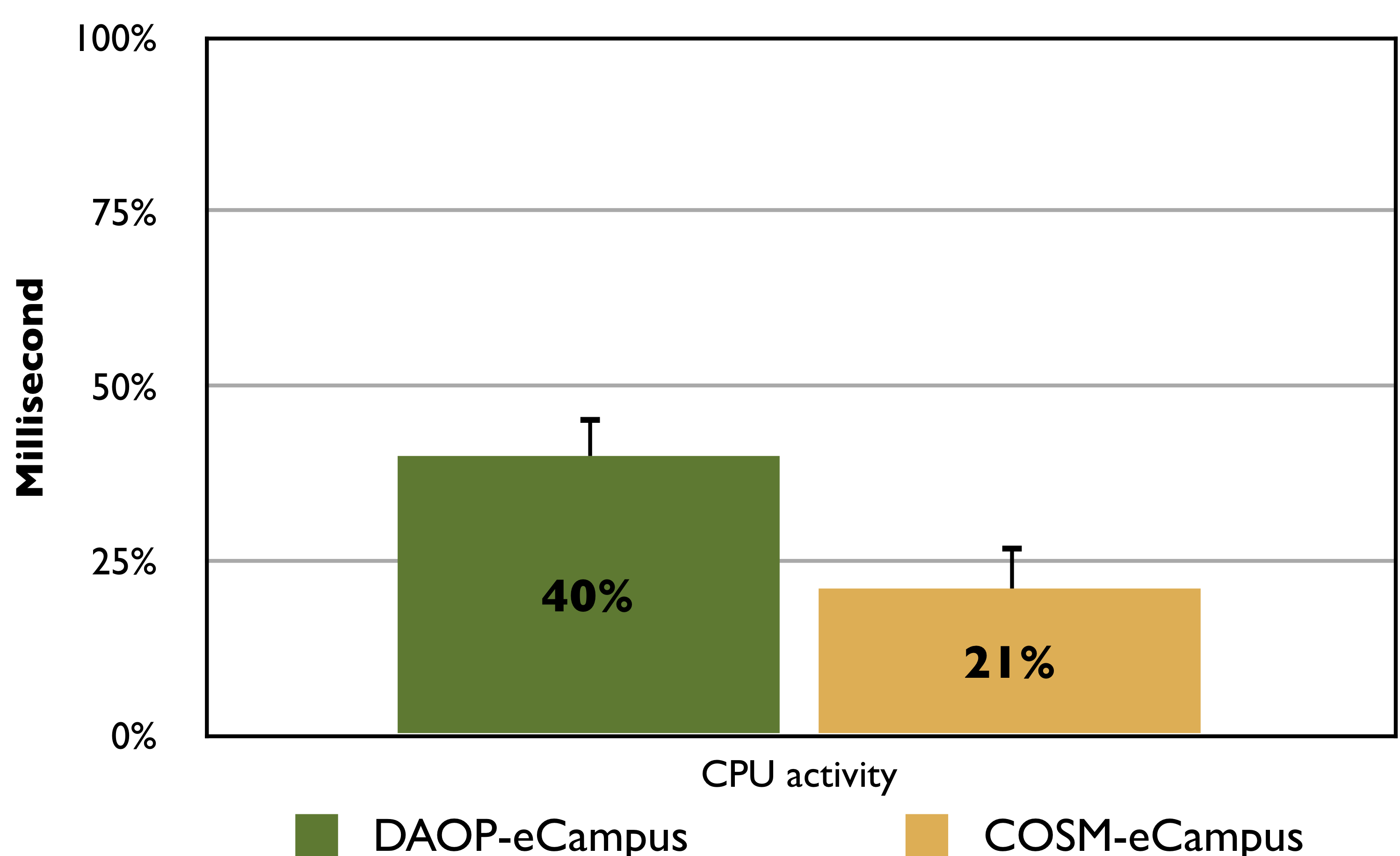}
  \caption{ Context Monitoring CPU Activity } 
\label{fig_CMCPU}
\end{figure}

\begin{figure}[!ht]
\centering
 \includegraphics[scale=0.54]{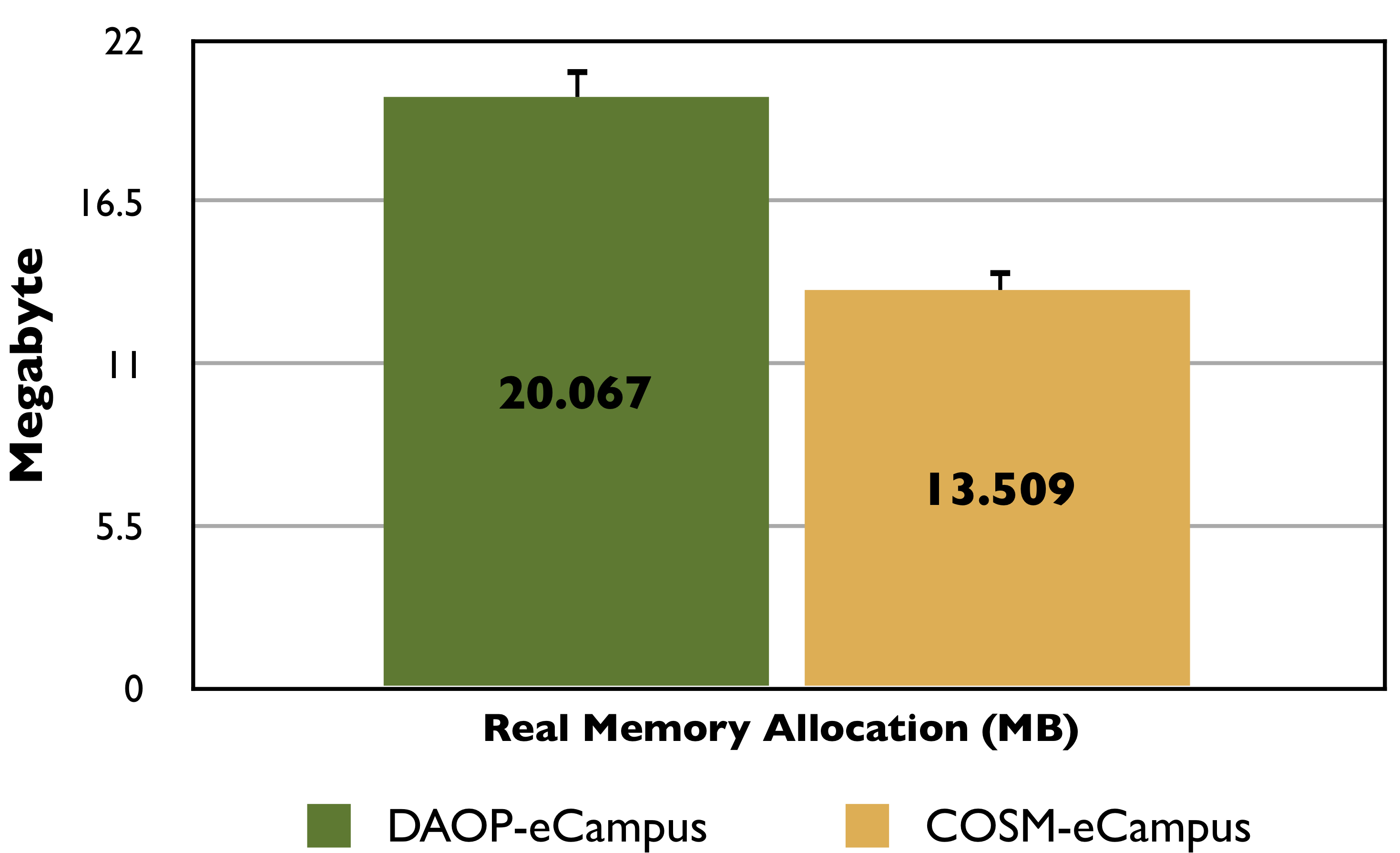}
  \caption{ Real Memory Allocation (MB) } 
\label{fig_CMmem}
\end{figure}

The results of the experiment on battery usage is shown in Figure \ref{fig_conexp}. This shows that the context-oriented component uses less battery energy than the DOAP implementation. The \gls{COSM}-middleware optimises the context-monitoring process by storing and processing the context information that was considered by the context-oriented component registration. Such enhancement of the context monitoring preserves the battery energy by 11.5\% of the total energy usage. This value is supported by the evaluation results of the CPU activity shown in Figure \ref{fig_CMCPU}. For context monitoring, the DOAP-eCampus requires more activity to be executed in comparison to COSD-eCampus; this consumes more battery energy. With regard to memory allocation, the DOAP-eCampus allocated more real memory to execute and perform the context snapshot (i.e. storing and processing the past context) than is needed by COSD-eCampus, as shown in Figure \ref{fig_CMmem}. This figure shows how expensive it is to allocate and process the context snapshot in the DOAP-eCampus application.

 \subsubsection{Experiment 2: Context Detection}
\begin{figure}[!ht]
\centering
 \includegraphics[scale=0.54]{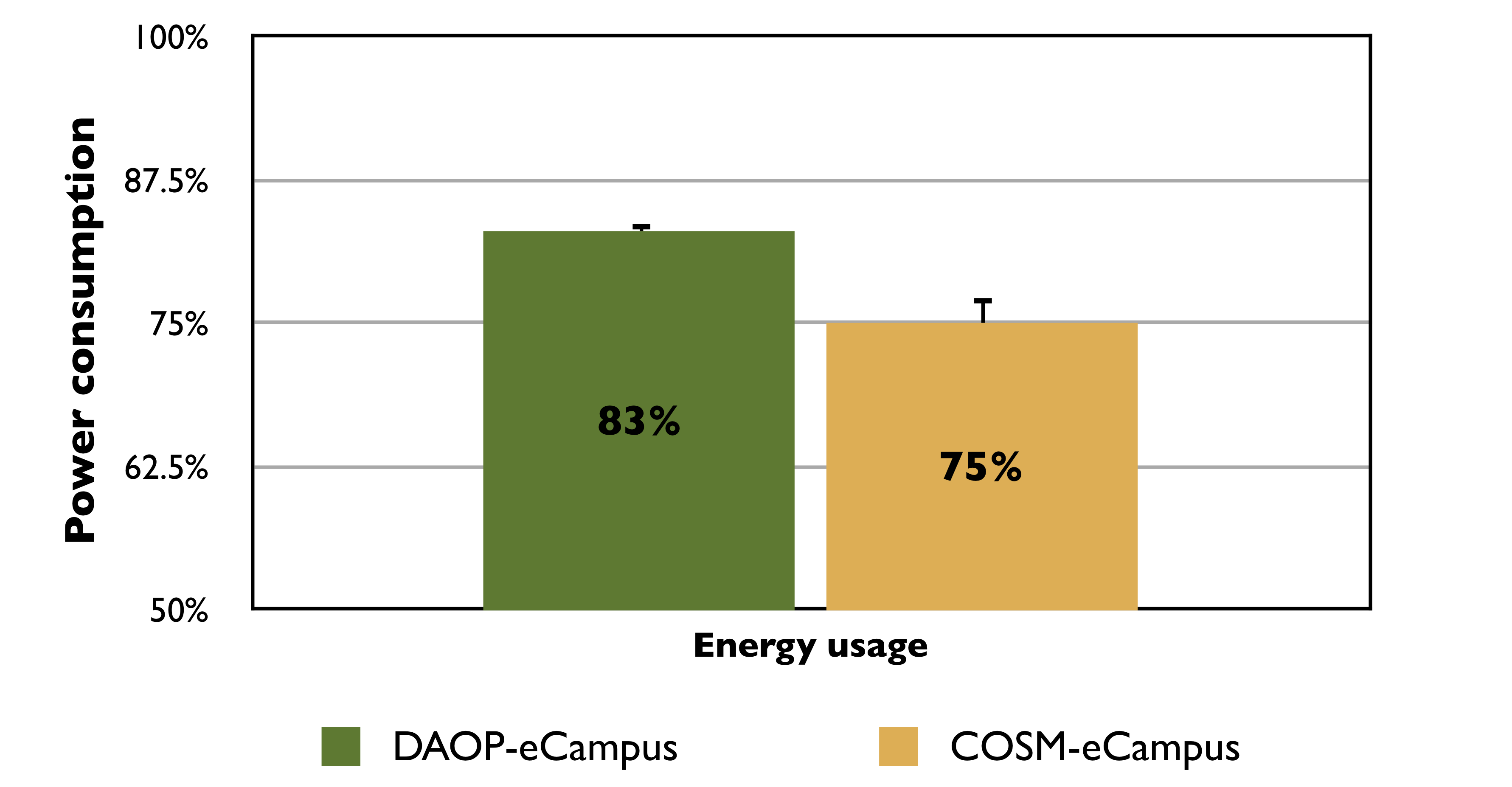}
  \caption{ Context Detection Battery Usage }
\label{fig_cdexp}
 \end{figure}

\begin{figure}[!ht]
\centering
 \includegraphics[scale=0.54]{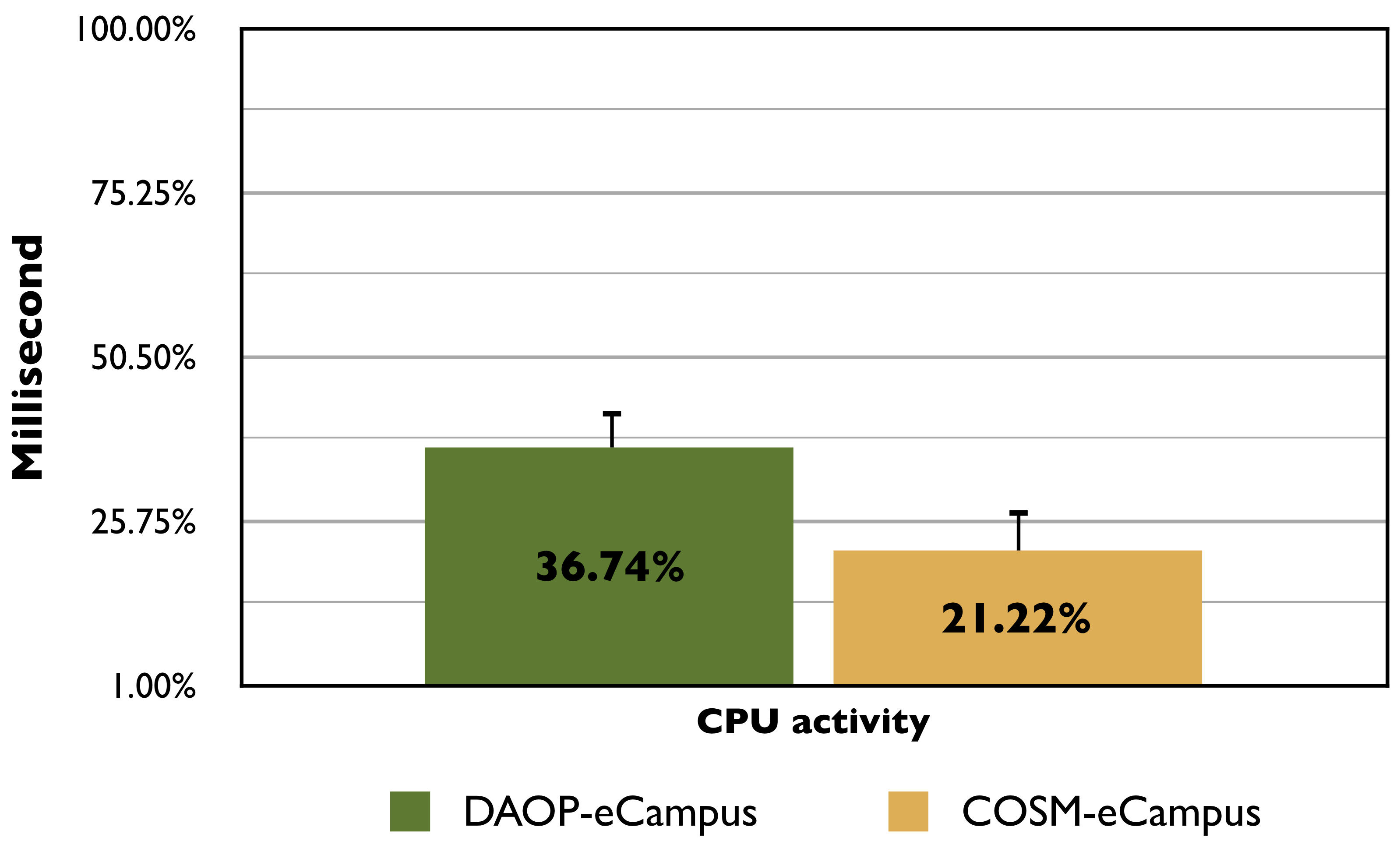}
  \caption{ Context Detection  CPU Activity }
\label{fig_CDCPU}
\end{figure}

\begin{figure}[!ht]
\centering
 \includegraphics[scale=0.54]{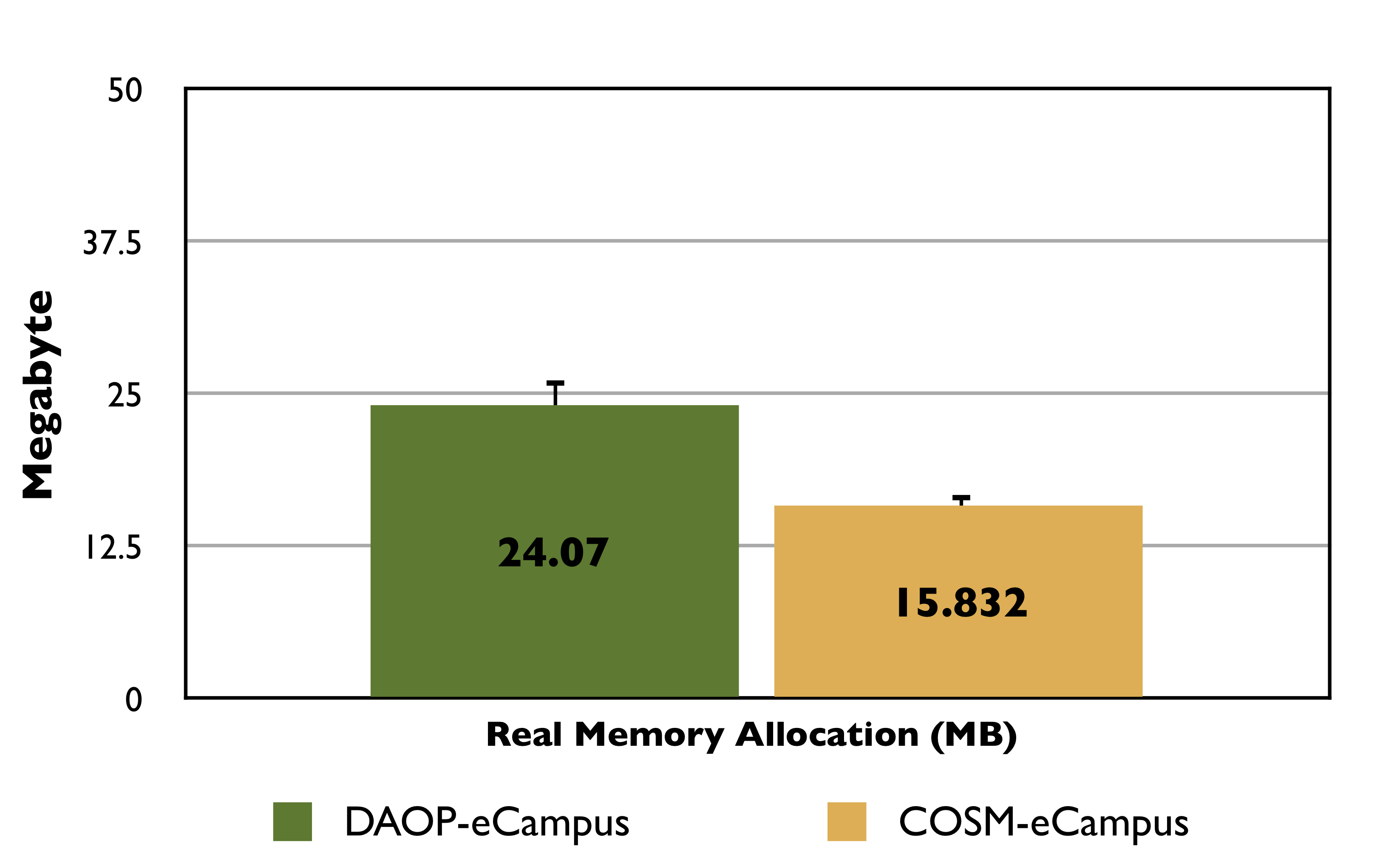}
  \caption{ Context Detection Real Memory Allocation (MB) }
\label{fig_Cdmem}
\end{figure}

For the context detection process, both implementations were evaluated based on the above criteria. The evaluation results for energy usage are shown in Figure \ref{fig_cdexp}. The evaluation results show that DOAP-eCampus consumes more energy to notify the application components about multiple context changes which were detected in short frequency. This requires more CPU activity to process the context changes and evaluate them with the passive context values stored in the joinpoints. The CPU activity for both applications is demonstrated in Figure \ref{fig_CDCPU}. In addition, the DOAP application requires more memory for allocating the aspect contexts and notifying them because each aspect must be allocated and executed. The \gls{AOP} framework then notifies the aspects about the context changes.  Later, the decision is left to the aspect methods implementation to decide whether to adapt or not. Such implementation of the context detection process using DOAP intensively consumes the allocated resources to notify multiple aspects about multiple events. In some cases, the aspect implementation was independent of the execution context, but it was executed and notified. The real memory allocation for the context detection process is shown in Figure \ref{fig_Cdmem}.   

  \subsubsection{Experiment 3: Activating Collaborative Aspects}

It is claimed that in \gls{AOSD}, dynamic aspect weaving can inject tangle-free code in the program execution; as explained before, context-dependent behaviours are collaborated aspects entangled with each other. It is claimed that in \gls{cosd}, context-oriented components can be activated dynamically to adjust the application behaviour, with affordable costs, during the adaptation. Designing context-dependent behaviour using an aspect-oriented programming paradigm requires platform support for activating aspects driven by the context state; such an implementation requires the \gls{AOP} platform to evaluate each joinpoint in conjunction with the associated context state and the passive context values. In addition, once the decision has been made, the \gls{AOP} platform must search for the associated method implementation that implements the required context-dependent behaviour. Moreover, from our own experience, it is very complex to decide which aspect should be woven first, because of the implicit dependence among the aspect implementations. For example, the platform should decide when the battery level is low, and which aspects must be activated. On the other hand, when activating the location aspect, the platform must consider the battery level before deciding which location service to use; such processes provide cyclic dependence among the aspects implementations and lead to unguaranteed adaptation outputs.

\begin{figure}[!ht]
\centering
 \includegraphics[scale=0.54]{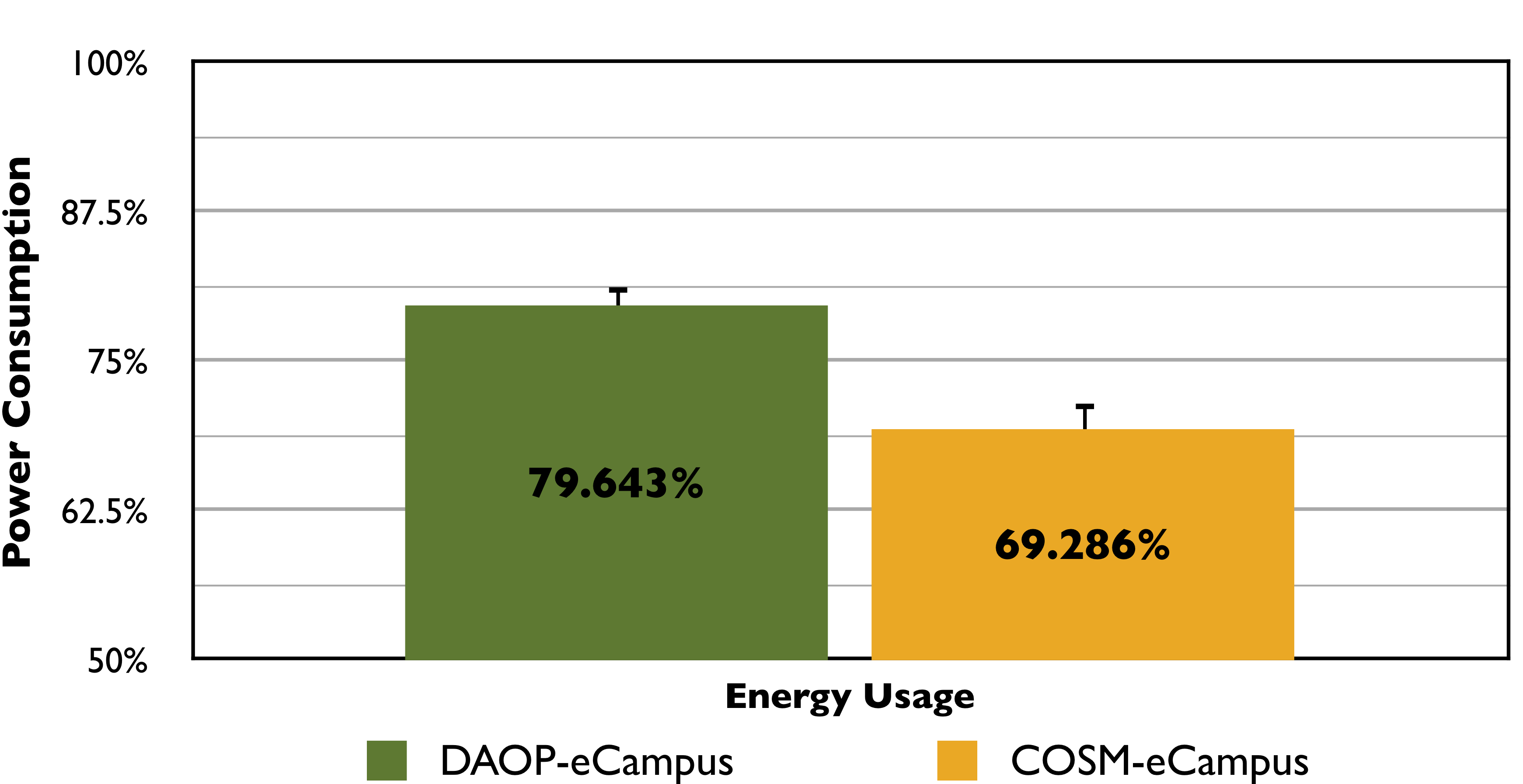}
  \caption{Activating Collaborated Aspects/context-oriented components Battery Usage }
\label{fig_caexp}
 \end{figure}

\begin{figure}[!ht]
\centering
 \includegraphics[scale=0.54]{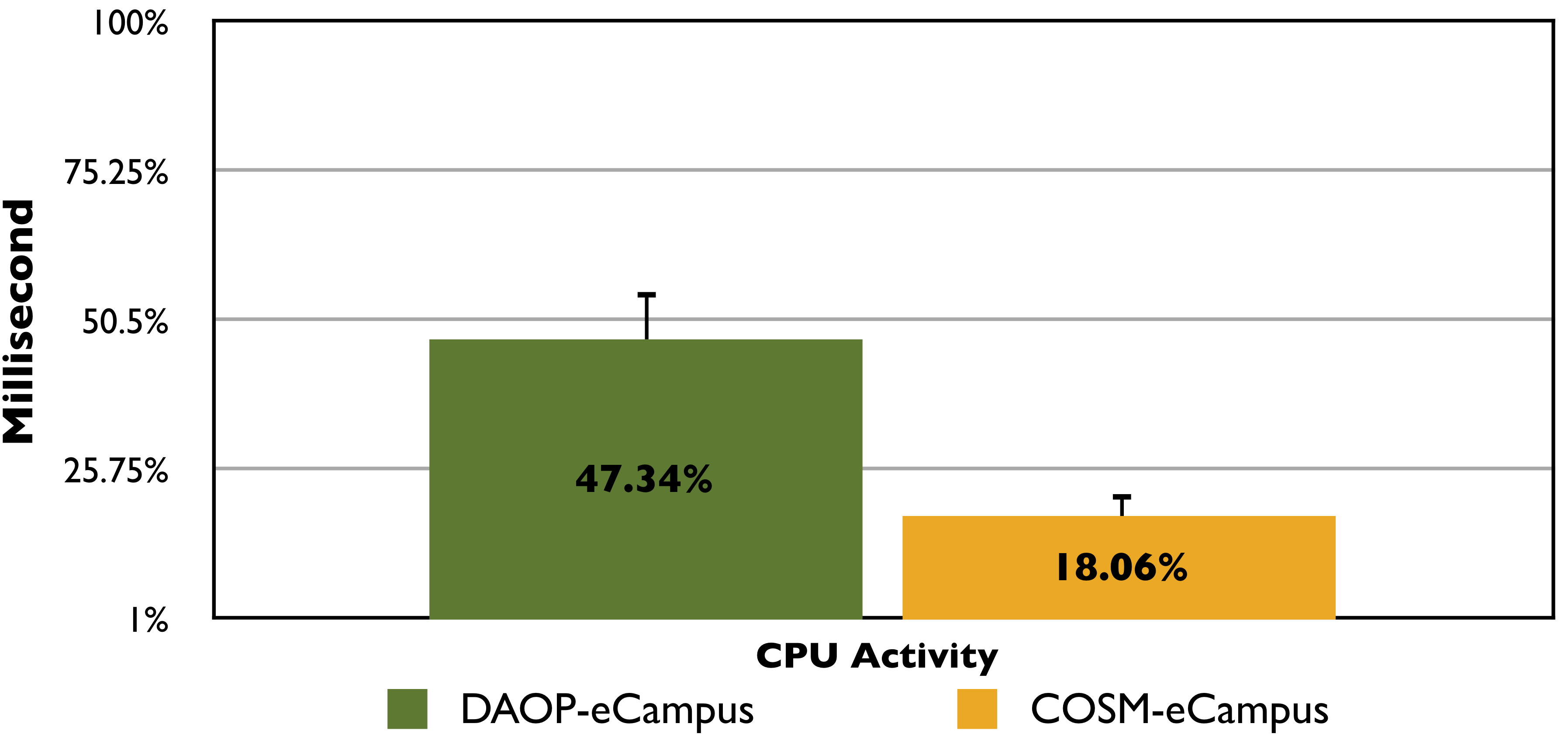}
  \caption{ Activating Collaborated Aspects/context-oriented components CPU Activity }
\label{fig_CACPU}
\end{figure}

\begin{figure}[!ht]
\centering
 \includegraphics[scale=0.54]{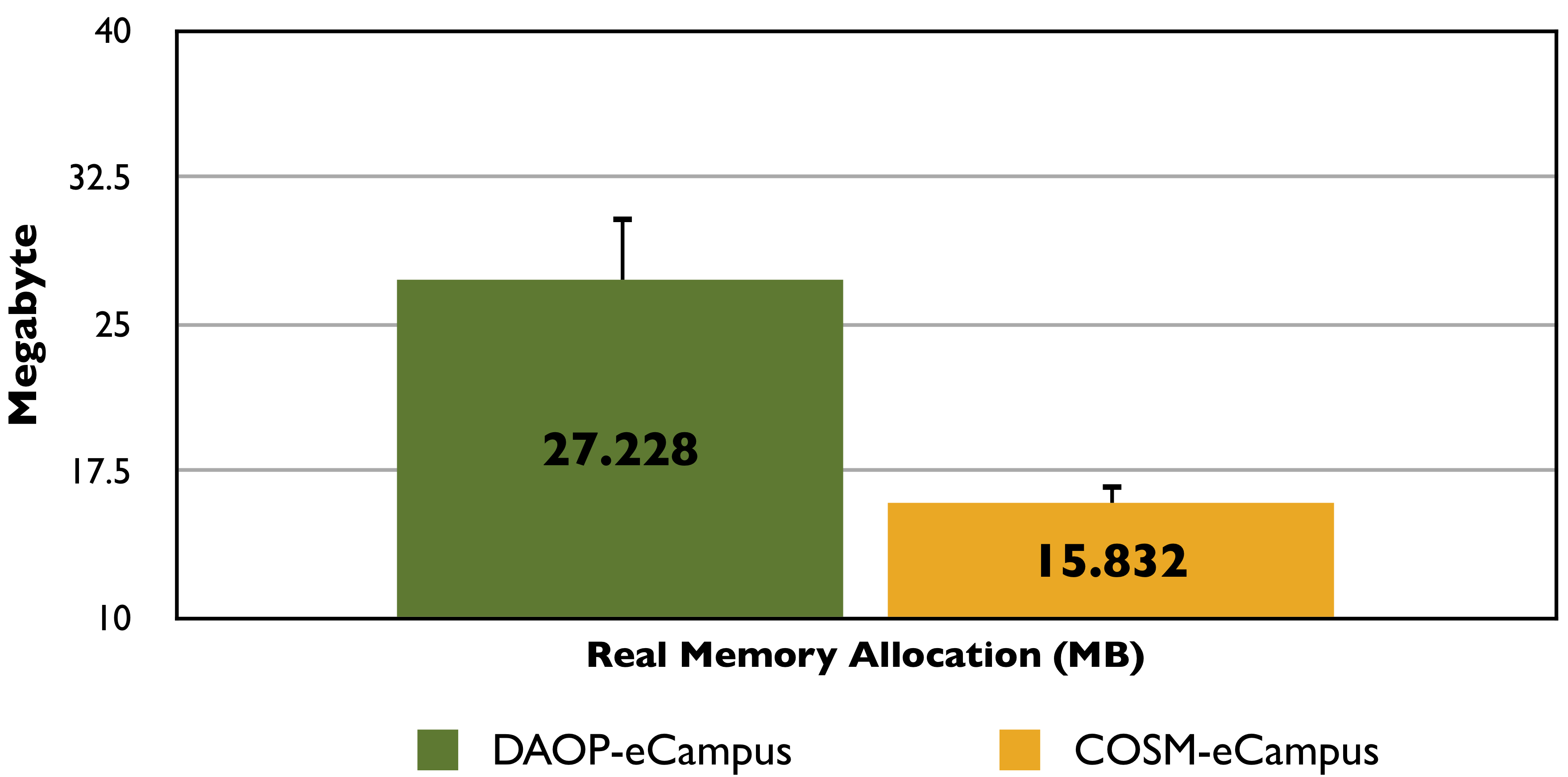}
  \caption{ Activating Collaborated Aspects/context-oriented components Real Memory Allocation (MB) }
\label{fig_CAmem}
\end{figure}

Figure \ref{fig_caexp} shows the battery usage when multiple contextual aspects are activated and executed compared to the composition of multiple context-oriented components. The figure shows that the \gls{DAOP}-eCampus consumes more energy to perform the adaptation, as it requires more energy to process the context state in each joinpoint. In addition, it requires the \gls{AOP} framework to resolve the dependence between several aspects before and after the \gls{Advice} methods execution. The CPU activity is shown in Figure \ref{fig_CACPU} and the real memory allocation for performing the activation and execution is shown in Figure \ref{fig_CAmem}.

\begin{figure}[!ht]
\centering
 \includegraphics[scale=0.54]{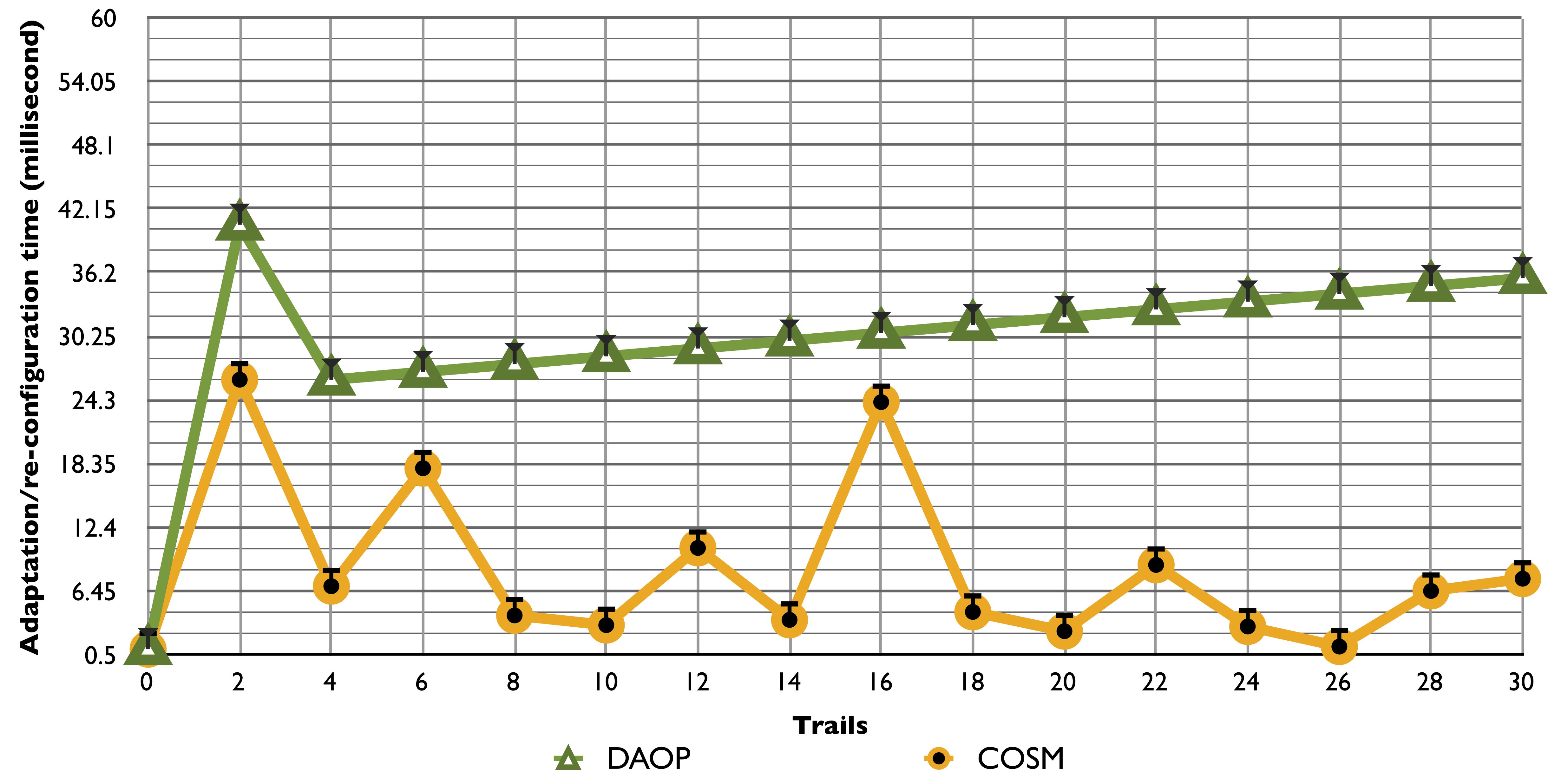}
  \caption{ Aspects/context-oriented components Composition}
\label{fig_caComp}
\end{figure}

The aspects composition needs to keep track of past context conditions and their associated states; more CPU activity and memory allocation are needed to perform this functionality. This experiment describes how each platform responds to multiple events (i.e. context conditions) detected at the same time. The adaptation/reconfiguration time for composing aspects/components is shown in Figure \ref{fig_caComp}. The values were taken every 2 min from the Apple Instruments tool while executing the application for 30 min continuously. As shown in Figure \ref{fig_caComp}, the COCA-eCampus requires less CPU time for composing the components, but DOAP requires more time for activating and executing the contextual aspects. The evaluation of aspects activation and execution shows an increased adaptation time because each aspect requires more memory allocation and CPU time to resolve the execution context with the context snapshot (i.e. context history). On the other hand, the \gls{COSM}-middleware requires more adaptation time for loading and executing the bundle implementation, but it can switch between weak/strong adaptation actions based on the execution context and the allocated resources. As shown in the figure, context-oriented components composition requires less adaptation/reconfiguration, based on the adaptation mechanism. Such variations in the adaptation time provided by \gls{COSM}-middleware can make use of the adaptation process and increase the device durability. The adaptation time in \gls{DAOP}, as shown in the figure, may increase over the execution time, which leads to poor performance and lower efficiency.

\subsection{Context-Oriented Middleware Evaluation\label{sec:mwev}}
The case study was implemented with \gls{COSM}-middleware and other approaches proposed in the literature. These approaches include the context-oriented programming paradigm targeting mobile devices, called \gls{JCOP}\cite{JCOL:2011p22222}, \gls{JCOOL}, supported by CAMEL methodology, which used aspect-oriented programming and middleware for context-dependent behaviours de/activation \cite{Sindico:2009p3478}, \gls{MUSIC}-middleware \cite{Geihs:2011p232}, and \gls{MADAM}-middleware \cite{Mikalsen:2006p4052}, which was fully implemented by Paspallis \cite{Paspallis:2009p3397}.
The implementation of the eCampus in \gls{JCOP} followed the COP approach \cite{JCOL:2011p22222}. The implementation in \gls{JCOOL} was accomplished with the aid of the aspect-oriented programming framework for Objective-C \cite{AspectCOCA:2011} and CAMEL framework. Both \gls{MUSIC}-middleware and \gls{MADAM}-middleware functionalities were implemented using the \gls{MUSIC} development paradigm proposed by Rouvoy et al.\cite{Rouvoy:2008p216}.

 \begin{figure}[!ht]
\centering
 \includegraphics[scale=0.54]{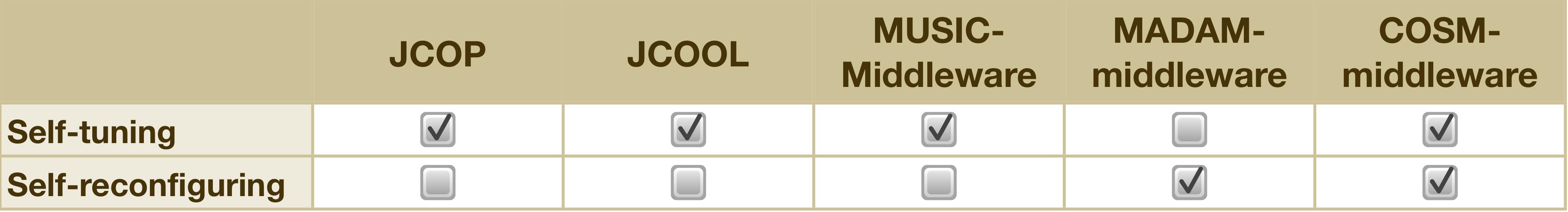}
  \caption{ Autonomic Properties Support }
\label{fig_selfstare}
\end{figure}
The objective of this experiment is to evaluate the ability of our \gls{COSM}-middleware in supporting the autonomic properties, self-tuning and self-configuring, of the self-adaptive eCampus application. The support of these properties in the above-mentioned solutions can be summarised as shown in Table \ref{fig_selfstare}. \gls{JCOP} and \gls{JCOOL} support only fine-grained adaptations using ad-hoc programming-level techniques; this is not able to change the application structure. \gls{MUSIC}-middleware supports both autonomic properties using parametric tuning and plug-in architecture. The implementation of plug-in architecture (i.e. building a software architecture from multiple plug-ins) used for adding or removing services/components at runtime. \gls{MADAM}-middleware supports only self-configuring property of self-adaptive system, as it models a separate plug-in architecture for each context provider. 
\subsection{Experiment 4: Self-tuning Evaluation}

\begin{figure}[!ht]
\centering
\includegraphics[scale=0.5]{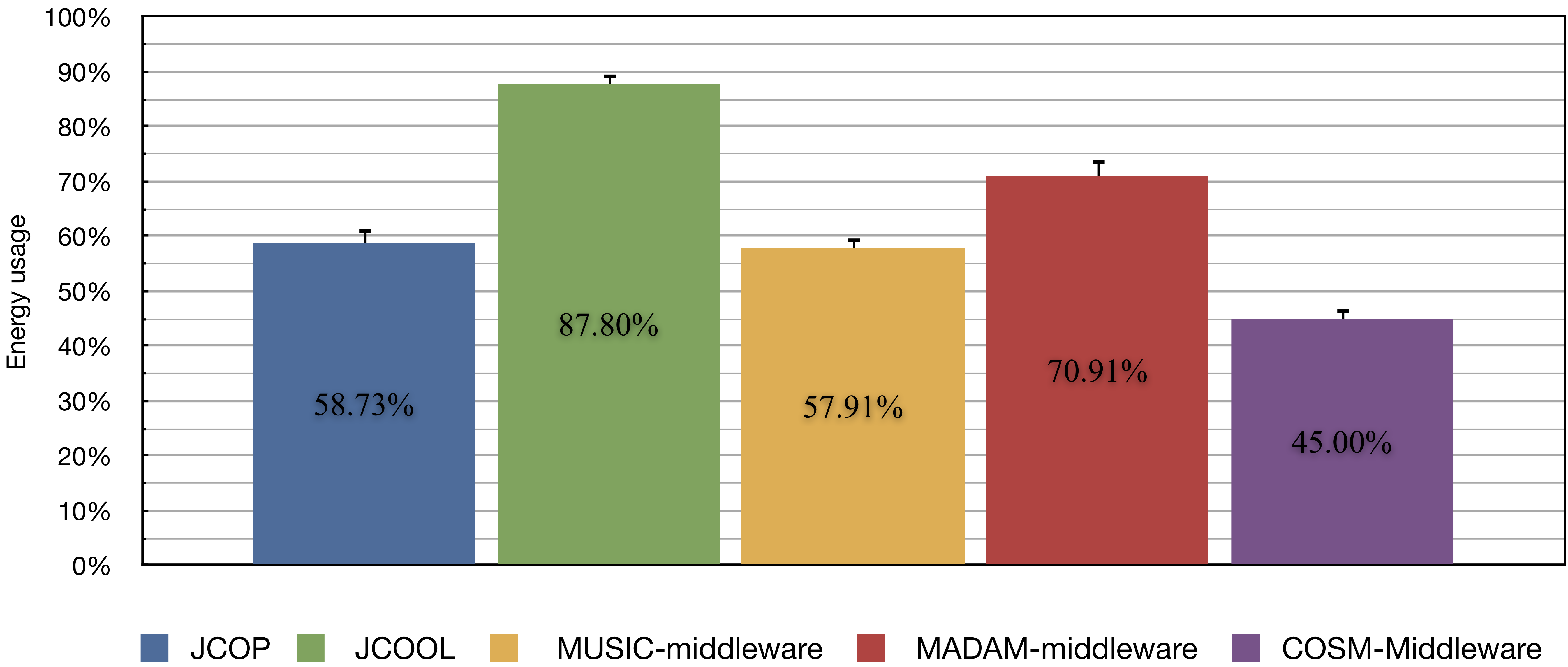}
 \caption{ Energy Usage for eCampus Application}
\label{fig_expr}
\end{figure}

Figure \ref{fig_expr} shows the experimental results for energy usage analysis for the eCampus running on the five platforms: \gls{JCOP}, \gls{JCOOL}, \gls{MUSIC}-middleware, \gls{MADAM}-middleware, and \gls{COSM}-middleware. The experiment shows that the COSD-eCampus implementation of the eCampus application used 15\% less battery energy than the \gls{MUSIC} implementation used. The eCampus implementation with \gls{JCOOL} consumes more energy during the adaptation processes because it does not consider the battery level or status during the adaptation action. In \gls{JCOOL}, the context values are only considered for evaluating the context-dependence in each joinpoint implementation. In the same way, the \gls{MADAM}-middleware drained the battery faster because each location service was implemented in a distinct plug-in (bundle) architecture, which requires more processing time for loading the bundle implementation. In contrast, when the same application was adapted by the \gls{COSM}-middleware, the application was able to adapt its behaviour and use less energy because the \gls{COSM}-middleware adapts to the location service by redirecting the delegate object to activate the required service implementation. In \gls{JCOP}, the application was able to adjust its behaviour and adapt the required location service; unfortunately, the context monitoring was relying on the mobile device operating system to deliver and detect the context information; this leads to faster consumption of the battery resource as a result of trying to process too many context events at the same time. The \gls{MUSIC}-middleware performs better with respect to battery consumption because \gls{MUSIC}-middleware uses a fine-tuning mechanism for manipulating components implementation. However, the application implemented using \gls{MUSIC}-middleware consumes more energy than that using \gls{COSM}-middleware because \gls{MUSIC}-middleware calculates the fitness of the application variant using a utility function every time the context state changes. Such verification at runtime requires more CPU time and memory allocation, which, as a result, consumes more energy. 

\begin{figure}[!ht]
 \includegraphics[scale=0.1]{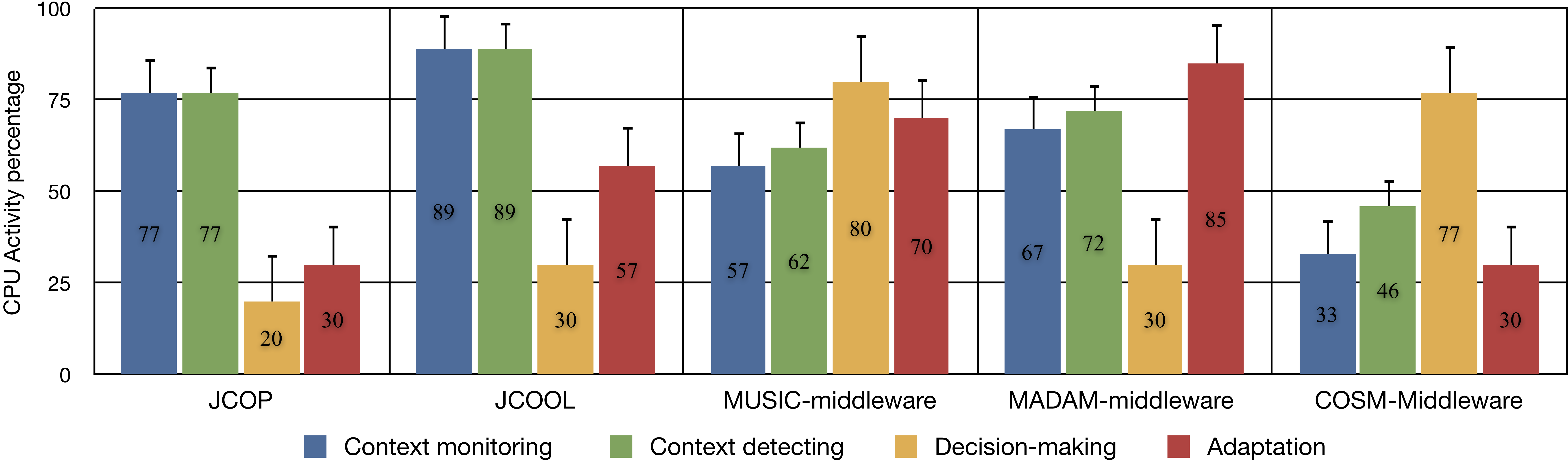}
  \caption{ Energy Usage for eCampus Application}
\label{fig_cpuact}
\end{figure}

Figure \ref{fig_cpuact} shows the experimental results for CPU activities analysed for the eCampus application in the five platforms previously mentioned. The evaluation considered adaptation processes including context monitoring, detecting, decision-making, and adaptation. As shown in Figure \ref{fig_cpuact}, context monitoring requires much more CPU activity in the \gls{JCOP} and \gls{JCOOL} platforms as they have no dedicated context-monitoring process, and they rely on the infrastructure to deliver the context information. The \gls{MUSIC} and \gls{MADAM} architectures come second with regard to context monitoring as they both implement a dedicated context manager which is able to process and filter the context information. Unfortunately, \gls{JCOP}, \gls{JCOOL}, \gls{MUSIC}, and \gls{MADAM} do not consider the effect of a contentious and unbalanced monitoring process for the context environment. This implies notifying the application several times about multiple context events, which requires more of the CPU time to process and handle these events. On the other hand, adapting the observer pattern allows the context manager in the \gls{COSM}-middleware to notify the interested components about the context changes when needed, so that the context-monitoring time drops from 57 ms in \gls{MUSIC}-middleware to 33 ms in \gls{COSM}-middleware. For the same reason, the context detection process drops from 72 ms to 46 ms in \gls{COSM}-middleware.

The decision-making in \gls{JCOP} and \gls{JCOOL} require less CPU activity as both platforms use static decision-making. The  \gls{JCOP} and \gls{JCOOL} approaches assume that the developers can predict when and where the context-dependent behaviour is needed in the application source code. For the same reason, the \gls{MADAM}-middleware requires less time for decision-making as it uses predefined rules supported by a rule engine to control the adaptation action. In the \gls{MUSIC}-middleware, the decision-making process is performed at runtime with the aid of a utility function; this requires more computation activities for analysing the architecture's constraints, adaptation goals, predefined rules, and the quality of services, plus the user's preferences. Afterwards, the application's variations model (adaptation plan) is selected based on the utility function results. In \gls{COSM}-middleware, the decision-making process is performed at runtime with the aid of the policy and verification manager, which both consider the decision policies in conjunction with the available resources and the quality-of-services (QOS). In general, the dynamic decision-making process requires more effort from the CPU than the static approach does, as shown in Figure \ref{fig_cpuact}.

 \begin{figure}[!ht]
\centering
\includegraphics[scale=0.5]{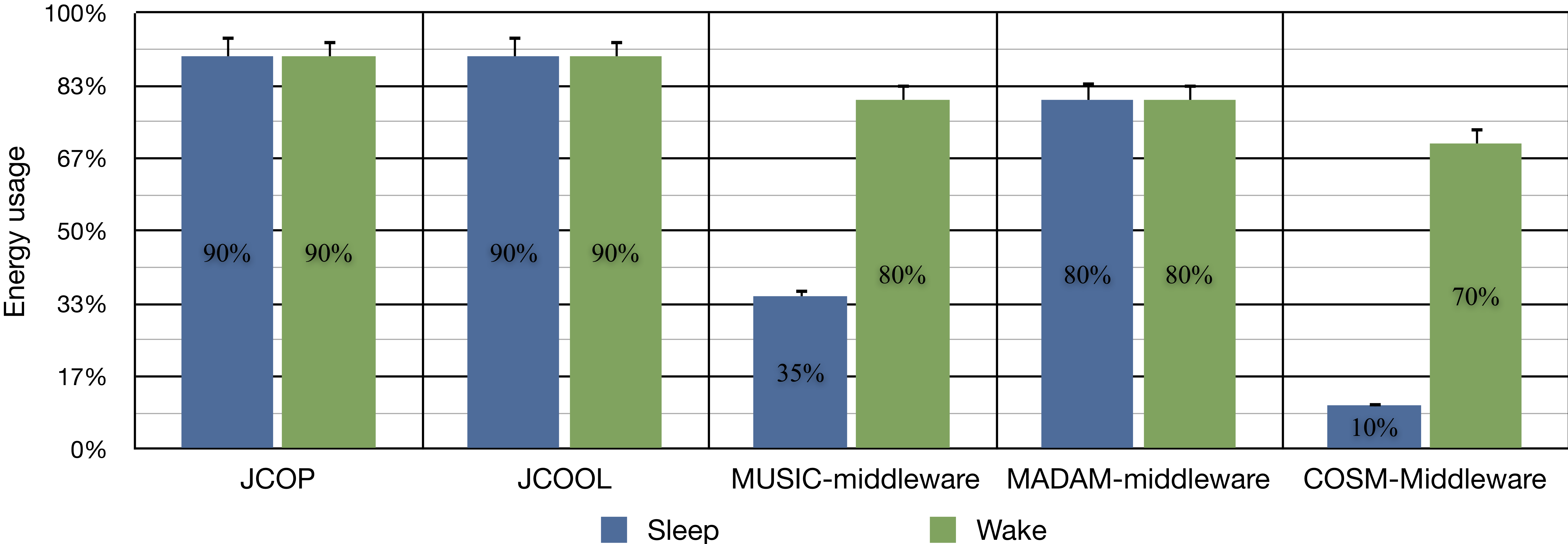}
 \caption{ Using the Allocated Resources in Sleep/Wake Mode}
\label{fig_sleep}
\end{figure}

Figure \ref{fig_sleep} shows how the five platforms use the allocated resources while the device is in Sleep/Wake mode. Such an evaluation reflects the middleware's ability to adjust its own functionality as long as the application is running in the background. At the same time, it shows how the \gls{COSM}-middleware wakens the application to notify the user about events in the monitored region. The other approaches have not considered the trade-off between the adaptation process and the allocated resources. The result of this is that the eCampus implementations on \gls{JCOP}, \gls{JCOOL}, and \gls{MADAM} were consuming battery life even when the application was in sleep mode or running in the background. The applications keep updating the current location of the device. Such an action was unnecessary, as the device was in sleep mode, so they consumed 90\% of the battery life after executing the application for 5 h. With the context-oriented midldeware and \gls{MUSIC}-middleware, the application was executed for the same period of time, and consumed less battery energy.
\subsection{Experiment 5: Self-reconfiguring Evaluation}
As mentioned before, self-reconfiguring is the capability of the software to adapt and behave autonomously in response to context changes. To evaluate this attribute, we considered the three versions: the \gls{MUSIC}-middleware, \gls{MADAM}-middleware, and \gls{COSM}-middleware. The \gls{JCOP} and \gls{JCOOL} platforms were excluded from this experiment as they did not support architecture reconfiguration.

\begin{figure}[!ht]
\centering
\includegraphics[scale=0.5]{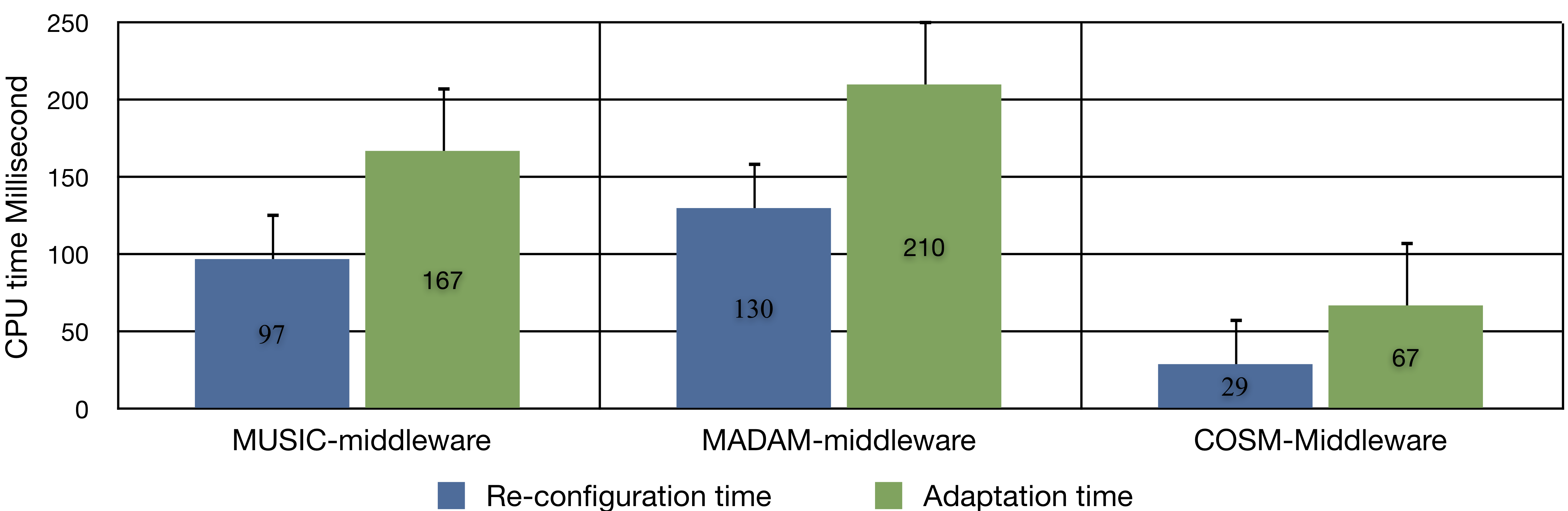}
\caption{ Adaptation/Reconfiguration Time (ms)  }
\label{table_adapy}
\end{figure}

\begin{figure}[!ht]
\centering
\includegraphics[scale=0.6]{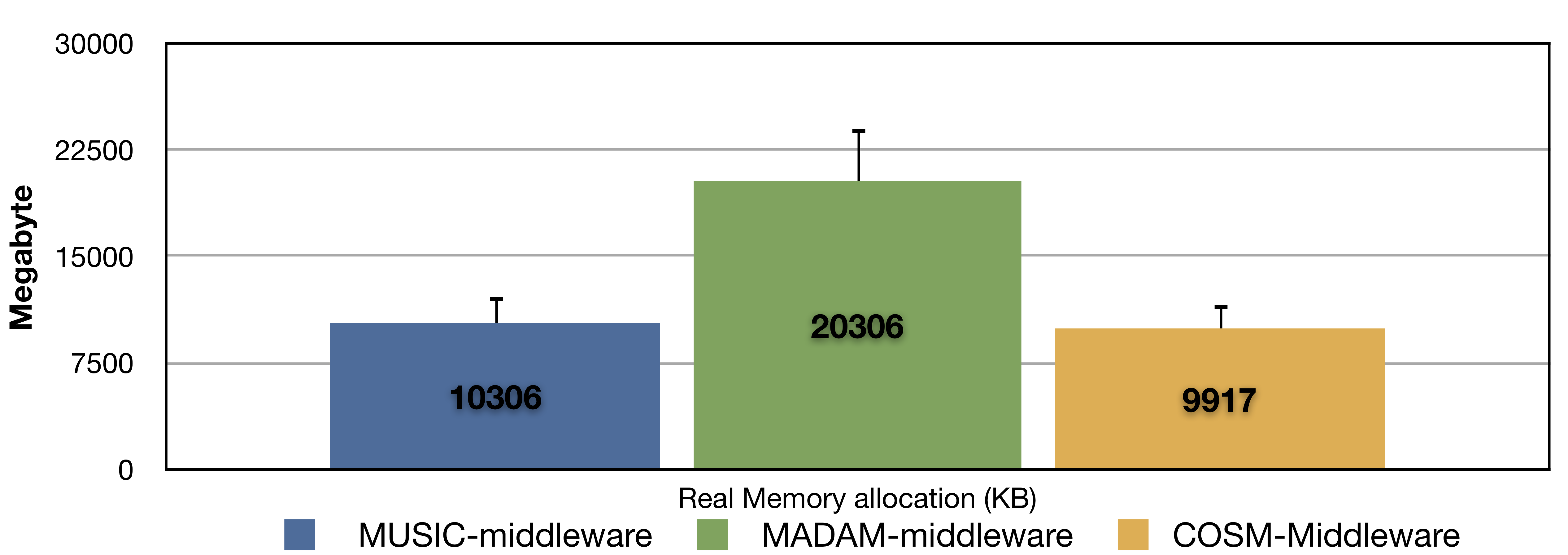}
\caption{ Memory Allocation for eCampus }
\label{table_mem}
\end{figure}

The adaptation/configuration time for adapting the eCampus application as mentioned before. The scenarios include adding a suitable location service according to the battery level, adding a location service component, and adapting to a web mapping service. Figure \ref{table_adapy} shows the evaluation results for the three architectures. The \gls{MADAM}-middleware requires more time to perform the adaptation because three bundles are loaded and executed. The total adaptation time was 210 ms. The \gls{MUSIC}-middleware required less time for reconfiguring the software as it adapted the location service using parametric tuning rather than loading a complete bundle for it. The \gls{COSM}-middleware comes first in the analysis, as it can switch autonomously between several location services by de/activating the associated layers. In addition, it verifies whether the plug-in can provide the necessary services before physically executing its implementation. In addition to this, according to the context state, the \gls{COSM}-middleware performs a runtime composition of the software components so that only the needed components are executed. In \gls{MUSIC}, the adaptation takes longer as it evaluates multiple application variations, then one variant is selected and executed. Loading a precompiled code from the component repository after instantiation (as done in the \gls{COSM}-middleware implementation) is accomplished in a shorter time than it takes to load the whole bundle implementation at once in the \gls{MUSIC}-middleware. This is illustrated in Figure \ref{table_mem}, which shows the memory allocations for the three platforms. It is worth mentioning here that when the battery level is low, the \gls{COSM}-middleware allocates less memory because of the size of the context-oriented component, which is small compared to the bundle implementation in the \gls{MUSIC}- and \gls{MADAM}-middlewares.

 \section{Related Work}
 \label{sec:RelatedWork}
 \glsresetall
Supporting the development and execution of self-adaptive software systems raises numerous challenges. These challenges include the development processes for building them, the design space, which describes the design patterns and the best practices of designing their building blocks, (i.e. component model or code fragments), and the adaptation mechanism, which describes the best adaptation action that can be used under the limited resources of the execution environment. The proposed approaches in the literature can be classified into model-centric, middleware-centric, and programming-level techniques. The ultimate goal of these approaches is to support adaptability, variability and increase the software quality by managing the context-dependent functionality at the programming level, middleware layer, or architecture model. In addition to that, they try to provide an adaptation mechanism, that have less impact on the allocated resources under the mobility constrains of the execution environments.

In self-adaptive applications, the selection of a particular component or code block at runtime is presumably made based on the active or passive context information plus the context state and its dependency \cite{Lincke:2010p4207}, and the possible composition of the context-dependent parts \cite{Hirschfeld:2008p1620}, which will exhibit volatile behaviour in the face of context changes. One could ask whether the current software domain techniques have sufficient support or suitable mechanism for performing such dynamic selection and composition of a software component based on its context-dependent functionality. Behavioural composition and re-configuration concerns require the software modules to be loosely coupled, and their behaviour variations can be combined and activated autonomously, according to the context changes. Such a challenge has been tackled by a composition strategy performed through the development process and depending totally on a static view of the self-adaptive software design. Such a view implies that the developers have to explicitly predict the final composition of the software and possible variations of the application using a programming-level technique such as \gls{COP} \cite{JCOL:2011p22222} and \gls{AOP} \cite{asosd:2004p1972}.
  

   \gls{COP} enables context-dependent adaptation and dynamic behaviour variations \cite{Gassanenko:1993p3934,Hirschfeld:2008p1620}. In \gls{COP}, context can be handled directly at the code level by enriching the business logic of an application with code fragments responsible for performing context manipulation, thus providing the application code with the required adaptive behaviour \cite{Salehie:2009p3693}. Costanza et al. \cite{Costanzza:2006p111} proposed the design of context-aware systems following a layered approach. The term " layer " refers to a specific context-dependent functionality, which might include a partial implementation of a class or a set of methods \cite{Costanzza:2006p111}. Alternatively, the whole class is encapsulated inside a layer \cite{Hirschfeld:2008p1620}. Hirschfeld et al. argued that the class-in-layer approach is more effective than the layer-in-class approach for encapsulating the context-dependent functionality, starting from the claim that context-dependent behavioural variations occur separately or in any combination, and in most cases they are collaborating and entangled with each other. A layer can be dynamically activated and composed with other layers, allowing fine-grained control of an application's runtime behaviour \cite{Hirschfeld:2008p1620}. An example for using \gls{COP} for implementing context-aware applications was proposed by Schuster et al. \cite{JCOL:2011p22222}. Schuster et al. proposed \gls{JCOP}, which uses a layered approach for achieving behavioural de-/activation for a prototype mobile application. The application was implemented based on a simple context model, which was implicitly encoded with the application code. Such approach shows the feasibility to use \gls{COP} for implementing self-tuning context-aware application for mobile computing environments. However, mobility induces \gls{Context} changes to the computational environment and therefore, changes to the availability of resources, and continuously evolving requirements would need software systems to be able to adapt to context changes. In most cases the developers on the provisional software design did not anticipate such context changes. Moreover, because of the software pervasiveness, and in order to make adaptation effective and successful, adaptation processes must be considered in conjunction with dependability and reliability by providing dynamic verification and validation mechanism, which validates the adaptation output with the adaptation goals, objectives, and architecture quality attributes \cite{Inverardi:2009p2345}. This requires the adaptation logic to be totally separated from the business code of the application. In addition, it needs a dynamic decision-making mechanism that maintains the architecture quality attributes during the adaptation.

For a more complex context-aware system, the same context information would be triggered in different parts of an application and would trigger the invocation of additional behaviours. In this way, context handling becomes a concern that spans several application units, essentially crosscutting into the main application execution. A programming paradigm aiming at handling such crosscutting concerns (referred to as aspects) is \gls{AOP} \cite{John:1997p4029}. \gls{DAOP} has emerged to enforce \gls{Separation of concerns} and support runtime adaptations through weaving code blocks in the application execution \cite{Popovici:2002p988}. The assumptions made by the \gls{COP} and \gls{AOP} approaches, i.e. that the developer knows all the possible software adaptations in advance and designs the application accordingly, is not sufficient to fulfil this need. In addition, in \gls{COP} and \gls{AOP} \cite{Tanter:2006p2222}, the context model and the adaptation logic are explicitly hard-coded in the application's business code \cite{Lincke:2010p4207}; this often leads to poor scalability and maintainability \cite{Kapitsaki:2009p3694}. In contrast, \gls{cosd}  separates the context model and the adaptation logic from the application code, which provides the software with the ability to adapt different context models at runtime without maintaining or modifying the application's business code \cite{magableh:2011p1231}.   

Dynamic weaving of aspects can be used for adjusting the software behaviour at runtime. However, existing \gls{DAOP} techniques tend to add a substantial overhead in both execution time and code size, which restricts their practicality for small devices with limited resources \cite{Hundt:2010p137}. The major reason for this poor performance is that the \gls{DAOP} architectures like PROSE 2 \cite{Popovici:2002p988} provide an AOP engine running at the Virtual Machine Layer. This engine accepts aspects at runtime, then transforms them into basic entities like \gls{Joinpoint} requests. Joinpoint refers to a point in the control flow of a program. In aspect-oriented programming a set of join points is described as a pointcut. A join point is a specification of when, in the corresponding main program, the aspect code should be executed. The \glspl{Joinpoint} are activated by registering them to the execution monitor. When the execution reaches one of the activated \gls{Joinpoint}, the execution monitor notifies the \gls{DAOP} engine, which executes the \gls{Advice} method after evaluating the actual and past activated contexts in each \gls{Joinpoint}. 
    
 The actual current solution in AOP frameworks is to take snapshots of context conditions only if necessary as stated in the Reflex framework \cite{Tanter:2006p2222}, or having a context repository to store the historical context information. However, context repository would not solve this problem because at each \gls{Joinpoint} the framework has to evaluate the actual (active) context with the passive (historical) context stored in the repository. This requires the framework to store and process the context history for multiple events at multiple times, which adds a substantial overhead to the allocated resources. In contrast, \gls{COSM}-middleware maintains and evaluates the architecture evolution by tuning the adaptation process. More specifically it adapts its own functionality and verifies the adaptation results dynamically by means of dynamic decision making.  Moreover, the \gls{COSM}-middleware reduce the tight coupling between the context providers and context consumers, it notifies the interested components about a specific context condition without any need to compare the current context condition with previous context values.

\gls{CAMEL} is an MDD-based approach proposed by Sindico and Grassi \cite{Sindico:2009p3478}. The approach uses a domain-specific language called \gls{JCOOL}, which provides a metamodel for context sensing with the supports of the context model designed using the \gls{JCOOL} meta model. However, Sindico and Grassi implemented the context binding as the associate relationship between context value and context entity. On the other hand, context-driven adaptation refers to structure or behaviour elements, which are able to modify the behaviour based on context values. The structural or behavioural insertion is accomplished whenever a context value changes; it uses \gls{AOP} inter-type deceleration, where the behavioural insertion is accomplished by means of an \gls{AOP} \gls{Advice} method to inject a specific code into a specific \gls{Joinpoint}. The authors used their former domain-specific language to support the \gls{COP} approach proposed by Hirschfeld et al. \cite{Hirschfeld:2008p1620}. Irrespective of this, \gls{JCOOL} is specific to an \gls{AOP} framework called the Simple Middleware Independent LayEr (SMILE) \cite{Smile:4432}. SMILE platform was used for distributed mobile applications \cite{Smile:4432}. The model approach in \gls{JCOOL} supports only ContextJ, which is an extension of the Java language proposed by Appeltauer et al. \cite{Appeltauer:2009p3541}. The \gls{CAMEL} methodology requires the software to be re-engineered whenever a new context provider is introduced into the context model. The developers must build a complete context model for the new values and maintain the underlying \gls{JCOOL} \gls{DSL} and the \gls{UML} model. The \gls{CAMEL} methodology has adapted \gls{AOP} and the \gls{EMF} to produce a context-oriented software similar to the layered approach proposed by Hirschfeld et al. \cite{Hirschfeld:2008p1620}. This makes \gls{CAMEL} limited to the \gls{EMF} tool support and the ContextJ language \cite{Haupt:2010p3399}. From our point of view \gls{CAMEL} tightly couples the software with modelling language, tool and the target deployment platform. In contrast,  \gls{COSM}-middleware is a generic adaptation engine provided by non-specialized language frameworks, and not being limited to a specific platform or mechanism. This gives the software developers the flexibility to construct a self-adaptive application using any object-oriented programming language and deploy it on several platforms \cite{magableh:2011p1231}.

  A context-driven adaptation requires the self-adaptive software to anticipate its context-dependent variations among its operational environment. The use of middleware in adapting the suitable adaptation approach provides a lead to achieving the adaptation results with less cost and several levels of granularity \cite{Salehie:2009p3693}. \gls{MADAM} aims to build adaptive applications for mobile devices using architecture models \cite{Floch:2006p2048}. The middleware is responsible for constructing and analysing several variability models at runtime, which adds an intensive overhead over the mobile device. \gls{MUSIC} middleware \cite{Rouvoy:2008p216} is an extension of the \gls{MADAM} component-based planning framework that optimises the overall utility of applications when context-related conditions occur. The planning-based adaptation of \gls{MADAM} employs dynamic configuration of component frameworks. In \gls{MUSIC}, the planning extends further, to support seamless configuration of component frameworks based on both local and remote components and services. Thus, both components and services are plugged in interchangeably to provide functionalities defined by the component framework. In MADAM and MUSIC the dynamic decision-making is supported by a utility function. A utility function is defined as the weighted sum of the different objectives based on user preferences and QoS. However, this approach suffers from a number of drawbacks. First, it is well known that correct identification of the weight of each goal is a major difficulty. Second, the approach hides conflicts among multiple goals in a single, aggregate objective function, rather than exposing the conflicts and reasoning about them. At runtime, a utility function is used to select the best application variant; this is the so-called 'adaptation plan'.

\section{Conclusions}
 In practice, performance and modifiability trade-off with each other as showed in the middleware evaluation. The \gls{COSM}-middleware achieves self-tuning and self-configuring without degrading the allocated resources. With regards to the adaptation processes including context monitoring, detecting, decision-making and adaptation, the \gls{COSM}-middleware shows better performance compared to other approaches proposed in the literature.  

The evaluation of the \gls{cosd} paradigm in comparison to \gls{AOSD} shows that using different decomposition mechanisms can affect the performance of aspects/components composition at runtime. The performance and energy usage in context-oriented applications are better than in \gls{DAOP}-applications. There is no doubt that Aspect-oriented frameworks can be used for developing and implementing self-adaptive applications, but their adaptability performance is poor in comparison to that of \gls{COSM}-middleware. The evaluation results show that implementing self-adaptive applications with the aid of \gls{COSM}-middleware can support software adaptability and variability with affordable adaptation costs and less impact on the allocated resources. Programming-level approaches like \gls{JCOP} and \gls{JCOOL} tend to support self-tuning of software systems with an acceptable level of performance, but the overall support for adaptability and variability is very limited in comparison with architecture evolution approaches such as \gls{MUSIC}, \gls{MADAM}, and \gls{COSM}-middleware. However, the programming techniques are better suited to small-scale context-dependent applications, and they require intensive modification for supporting context monitoring, context detection, and dynamic decision-making.

\bibliographystyle{IEEEtran}

\bibliography{references.bib}
 
\end{document}